\documentclass[draft]{agujournal2019}
\usepackage{url} %this package should fix any errors with URLs in refs.
\usepackage{lineno}
\usepackage[finalnew]{trackchanges} %for better track changes. finalnew 
\tcregister{\cite}{7}
\tcregister{\ref}{7}
\usepackage{soul}

\usepackage{graphicx}
\usepackage{appendix}
\usepackage{float}
\usepackage{xcolor}
\usepackage{multirow}
\usepackage{amsmath, amsfonts, amssymb, mathtools}
\usepackage[T1]{fontenc}
\usepackage{newtxtext,newtxmath}
\usepackage{bm}
\usepackage{multicol}
\usepackage[font=footnotesize]{caption}
\usepackage{csquotes}
\usepackage{siunitx}
\usepackage{array} 
\usepackage{makecell}

\addeditor{R1}
\addeditor{R2}
\addeditor{a}

\draftfalse

\journalname{JGR: Space Physics}

\begin{document}
\makeatletter
\def\@oddhead{\hfil{\footnotesize\itshape\color{gray} In press in JGR: Space Physics}\hfil}
\let\@evenhead\@oddhead
\makeatother

%%%%%%%%%%%%%%%%%%%%%%%%%%%%%%%%%%%%%%%%%%%%%%%
%  TITLE
%
% (A title should be specific, informative, and brief. Use
% abbreviations only if they are defined in the abstract. Titles that
% start with general keywords then specific terms are optimized in
% searches)
%
%%%%%%%%%%%%%%%%%%%%%%%%%%%%%%%%%%%%%%%%%%%%%%%

% Example: \title{This is a test title}

\title{High-Precision Long-Time Numerical Solutions of Charged-Particle Motion in a Static Magnetic Dipole Field}

\authors{H.L. Jiles\affil{1} and R.S. Weigel\affil{1}}

\author{R.S. Weigel}
\affil{}
\date{\today}

\affiliation{1}{Space Weather Lab\\Department of Physics and Astronomy\\George Mason University}

\correspondingauthor{H.L. Jiles}{hjiles@gmu.edu}

%%%%%%%%%%%%%%%%%%%%%%%%%%%%%%%%%%%%%%%%%%%%%%%
% KEY POINTS
%%%%%%%%%%%%%%%%%%%%%%%%%%%%%%%%%%%%%%%%%%%%%%%
%  List up to three key points (at least one is required)
%  Key Points summarize the main points and conclusions of the article
%  Each must be 140 characters or fewer with no special characters or punctuation and must be complete sentences

% Example:
% \begin{keypoints}
% \item	List up to three key points (at least one is required)
% \item	Key Points summarize the main points and conclusions of the article
% \item	Each must be 140 characters or fewer with no special characters or punctuation and must be complete sentences
% \end{keypoints}

\begin{keypoints}
\item Relativistic particle trajectories in a dipole magnetic field are simulated for up to $\sim\!10$ years with the PS method and compared with RK methods.
\item Relative errors in constants of motion, kinetic energy and canonical momentum, are below $10^{-10}$ for the PS method over $\sim\!10^5$ gyroperiods.
\item The PS method is shown to have significant advantages in speed and accuracy over several RK-based methods.
\end{keypoints}

%%%%%%%%%%%%%%%%%%%%%%%%%%%%%%%%%%%%%%%%%%%%%%%
%
%  ABSTRACT and PLAIN LANGUAGE SUMMARY
%
% A good Abstract will begin with a short description of the problem
% being addressed, briefly describe the new data or analyses, then
% briefly states the main conclusion(s) and how they are supported and
% uncertainties.

% The Plain Language Summary should be written for a broad audience,
% including journalists and the science-interested public, that will not have 
% a background in your field.
%
% A Plain Language Summary is required in GRL, JGR: Planets, JGR: Biogeosciences,
% JGR: Oceans, G-Cubed, Reviews of Geophysics, and JAMES.
% see http://sharingscience.agu.org/creating-plain-language-summary/)
%
%%%%%%%%%%%%%%%%%%%%%%%%%%%%%%%%%%%%%%%%%%%%%%%

%% \begin{abstract} starts the second page

\begin{abstract}
The motion of a charged particle in a magnetic dipole field has no analytical solution, but has two exact invariants: kinetic energy and canonical azimuthal angular momentum. The guiding center approximation is frequently used to understand trapped particle motion, but is valid only for limited times, and its assumptions are not satisfied for all trapped particles. Long-time simulations have historically been limited by computational speed and numerical accuracy. We investigate the Parker--Sochacki power series (PS) method, comparing it with Runge--Kutta (RK) methods and using exact invariant conservation to measure accuracy. Unlike the fourth-order (RK4), Dormand--Prince (RK45), and symplectic Gauss--Legendre (RKG) methods, which use derivative estimates at each step, PS uses a power series expansion in time. We compare methods using identical fixed time steps for protons and electrons across multiple energies and pitch angles, over durations up to ${\sim}10^9$ gyroperiods (${\sim}10$ years). \change[R2]{PS conserves the kinetic energy 4 to 13 orders of magnitude better than the RK methods, and the canonical momentum 4 to 10 orders of magnitude better than RK4 and RK45 (comparably to RKG).}{PS conserves kinetic energy at least 10 orders of magnitude better than RK4 and 4 to 7 orders better than RKG; canonical momentum, 4 to 11 orders better than RK4 and comparably to RKG. It matches RK45's strictest-tolerance accuracy in under a tenth of the runtime.} \change[R2]{When compared using a target relative kinetic energy error, PS is approximately ten times faster than RK4 and twice as fast as RKG in representative runs.}{At a target accuracy, PS has the lowest errors while being several times faster than the explicit RK methods and comparable to RKG.} \change[a]{Results demonstrate that the PS method can be used to evaluate}{The PS method can therefore be used to evaluate} the long-time behavior of charged particles trapped in a dipole field, over timescales where the validity of the guiding center approximation is unknown and for initial conditions violating its assumptions.

%PS is the only method that is accurate and stable across all simulations considered in this study.
%In long-time simulations, PS errors grow approximately linearly over $10^9$ equatorial gyroperiods, whereas RKG develops secular energy error growth beyond $10^6$ equatorial gyroperiods.
\end{abstract}

%\section*{Plain Language Summary}
%Enter your Plain Language Summary here or delete this section.

%https://www.agu.org/Share-and-Advocate/Share/Community/Plain-language-summary

%%%%%%%%%%%%%%%%%%%%%%%%%%%%%%%%%%%%%%%%%%%%%%%
%
%  BODY TEXT
%
%%%%%%%%%%%%%%%%%%%%%%%%%%%%%%%%%%%%%%%%%%%%%%%

%%% Suggested section heads:
% \section{Introduction}
%
% The main text should start with an introduction. Except for short
% manuscripts (such as comments and replies), the text should be divided
% into sections, each with its own heading.

% Headings should be sentence fragments and do not begin with a
% lowercase letter or number. Examples of good headings are:

% \section{Materials and Methods}
% Here is text on Materials and Methods.
%
% \subsection{A descriptive heading about methods}
% More about Methods.
%
% \section{Data} (Or section title might be a descriptive heading about data)
%
% \section{Results} (Or section title might be a descriptive heading about the
% results)
%
% \section{Conclusions}

\section{Introduction}\label{SEC:intro}

Many physical systems can be modeled by ordinary differential equations (ODEs), yet closed-form solutions exist only for linear problems and a limited set of specially structured nonlinear problems, making numerical methods essential \cite{Chapra2021}. Space physics offers a demanding example in which the motion of a charged particle in a magnetic dipole field has no general analytical solution \cite{DragtFinn1976} yet has two exact invariants of motion: kinetic energy and canonical azimuthal angular momentum (hereafter referred to as canonical momentum). Numerical solutions of the long-time dynamics must conserve these invariants across many gyroperiods, posing an accuracy challenge. 

%Chapra2021: 
%"For the time being, the important fact is that exact solutions for many ODEs of practical importance are not available. As is true for most situations discussed in other parts of this book, numerical methods offer the only viable alternative for these cases."
%"One very important method that engineers and applied mathematicians developed to overcome this dilemma was linearization."
%"Although linearization remains a very valuable tool for engineering problem solving, there are cases where it cannot be invoked."
%"However, as described in the previous section, many of the differential equations of practical significance cannot be solved using the analytical methods of calculus. Thus, the methods discussed in the following chapters are extremely important in all fields of engineering"
%"The most popular RK methods are fourth order. As with the second-order approaches, there are an infinite number of versions."

A particle in a dipole magnetic field with velocity components both parallel and perpendicular to the magnetic field will undergo multiple simultaneous periodic motions with respect to field lines: gyration around, oscillation (bounce) along, and drift across \cite{Northrop1963, Dragt1965, Walt1994}. The adiabatic invariants are the basis for the guiding center approximation, which is valid when the gyroradius, $r_g$, remains small compared to the length scale over which the magnetic field varies, $L_B \equiv {B}/{|\nabla B|}$, such that $\varepsilon \equiv r_g/L_B \ll 1$ \cite{Northrop1963}. Their long-term behavior is central to understanding trapped-particle dynamics over timescales that are long compared to the gyroperiod \cite{DragtFinn1976}.

Although the guiding center approximation and the assumption of adiabatic invariance are sufficient for addressing many problems in trapped-particle dynamics, not all problems of interest satisfy this assumption, in which case numerical integration of the exact equations of motion is required. Examples include the tracing of high-energy protons and electrons in radiation belts, where the gyroradius is not small compared with the scale over which the field varies \cite{Yugo2001, Yugo2007}, and the chaotic scattering of relativistic protons in the inner magnetosphere, where the motion is nonadiabatic, and their trajectories have a sensitive dependence on initial conditions \cite{Meseguer2025}. Similarly, \citeA{Oran2022} found that trapped-particle trajectories around small magnetized bodies and planets become chaotic once the gyroradius exceeds a fraction of the field line radius of curvature and, because of the sensitivity to initial conditions, cannot be reliably integrated on standard computers due to roundoff errors \cite{LiLiao2018}. Another historical example is the long-time conservation of the magnetic moment and the implications for the guiding center approximation. \citeA{DragtFinn1976} concluded that the use of adiabatic invariants to predict the long-time behavior of radiation belt particles has no rigorous justification, mathematical or numerical, and expected roundoff errors to defeat any numerical test. Whether the magnetic moment remains effectively conserved over long timescales is an open question. \add[R1]{More recent studies have shown that, depending on the energy and pitch angle, trapped particles may show nonadiabatic magnetic moment behavior and chaotic orbits \cite{Murakami1990, Xie2020, Oran2022, Meseguer2025}.}

Addressing such problems numerically places substantial demands on the integrator. In each example, trajectories must remain accurate over many gyroperiods so that numerical error is not mistaken for physical effects.

Runge--Kutta (RK) based numerical ODE solution methods have become standard tools \cite{Chapra2021, Butcher2016}; they advance the solution incrementally by sampling derivative estimates at each time step. To achieve high accuracy, problem-specific modifications to the RK algorithm are needed. \citeA{Yugo2001, Yugo2007} used a symplectic Gauss--Legendre Runge--Kutta method and \citeA{Meseguer2025} used a ninth-order Runge--Kutta method with an adaptation from \citeA{Calvo2006} that enforces the conservation of the invariants at each step. In both cases, maintaining low error in the exact invariants required modifications to the RK algorithm. 

%that resulted in bounded the energy errors over $10^7$ steps.
%and reported conservation of energy and canonical momentum to near machine precision over approximately $10^2$ bounces.

%Calvo2006: ""A new strategy to preserve invariants in the numerical integration of initial value problems with explicit Runge–Kutta methods is presented. It is proved that this technique retains the order of the original method, has an easy and cheap implementation, and can be used in adaptive Runge–Kutta codes.....the above projection techniques can be included in the general form ϕ̃_h(y_n) = ϕ_h(y_n) − λ_n Φ_n, where the vector Φ_n ∈ ℝ^m defines the direction of projection and λ_n is a scalar chosen so that ϕ̃_h(y_n) belongs to the invariant manifold." The project aspect is very abstract to me but I took this to mean that it enforces the conservation at each step...Meseguer is a tad bit more clear: "To preserve conserved quantities, this time-stepper has been suitably modified by adding a projection of the predicted trajectory over the invariant manifold, as described in Refs. [19–21]." (Meseguer et al. 2025, Appendix Sec. IV)." I understood this to mean that they take the RK step, then move the result back onto the surface where energy and P_φ have their initial values.

In this work, we implement the Parker--Sochacki (PS) power series method for relativistic charged-particle dynamics in a static magnetic dipole field and evaluate its ability to conserve the system's exact invariants. In a subsequent work, we will address applications of high-accuracy long-time solutions that require the understanding of the achievable accuracy documented here. The PS method is a conceptually distinct alternative to RK-based methods: rather than estimating derivatives at each time step, it uses a time-dependent power-series expansion of the solution specific to the system of equations. In contrast to RK-based methods, achieving high-accuracy solutions for the dipole problem requires no problem-specific modifications to the PS algorithm. However, the equations to solve are problem-dependent because the power series depends on the equations of motion.

The remainder of this paper is organized as follows. In Section~\ref{SEC:numerical methods}, we describe the numerical methods compared in this work and the criteria of accuracy, stability, and computational efficiency used to assess them. In Section~\ref{sect:implementation}, we provide details on how the ODE solvers are implemented. In Section~\ref{sec:dipole}, we derive the PS recurrence relations for a charged particle in a static magnetic dipole field. In Section~\ref{SEC:results}, we benchmark the methods, first using a common fixed time step and then using a minimum target accuracy. From these results, we demonstrate that the PS method conserves the exact invariants of the static magnetic dipole problem, the kinetic energy and the canonical momentum, \change[R2]{with higher precision than both conventional and symplectic RK methods, without auxiliary constraints and with similar or higher computational efficiency}{to high precision without modifications to impose conservation. It gives smaller errors than the conventional and symplectic RK methods, with comparable canonical momentum errors to the symplectic method, and matches the precision of a tightly toleranced adaptive RK45 method at a fraction of the runtime}. Section~\ref{SEC:Conclusions} summarizes our conclusions.

%Dragt1965:"The numerical results of the last section show that α and therefore J are remarkably constant from one equatorial crossing to the next. But this does not really answer the important question: what happens to J after many equatorial crossings?" Note that J is the "gyro-action"....some difficult-to-follow terminology in the '65 paper.

%DragtFinn1976:"(the first term in the magnetic moment series and the expression usually assumed to be constant in practical calculations) typically changes by 1%–50% between equatorial crossings. The truncated magnetic moment series Γ, with a truncation chosen to minimize variations, typically varies by 1%–0.01%."
%"accumulate neither additively nor randomly but instead must be highly self cancelling."

\section{Numerical Methods}\label{SEC:numerical methods}

In standard RK methods, derivative estimates are sampled at selected points, called stages, which are combined into a weighted average to estimate the increment in the solution at each time step. The number of stages and the chosen weights determine the order of the method. While many variations of this method exist, the fourth-order Runge--Kutta (RK4) method is the most widely used \cite{Chapra2021, Walters2022} because it provides a good balance between accuracy and computational runtime. Because RK methods approximate the solution incrementally from local derivative estimates, their step sizes are constrained by numerical stability and accuracy requirements, particularly in stiff or rapidly varying systems, where very small time steps may be necessary. For such problems, the adaptive Runge--Kutta Dormand--Prince (RK45) method, which adjusts the time step size based on a local error estimate, can be used \cite{Dormand1980,Chapra2021}. For problems where long-time stability or the conservation of exact invariants (constants of motion) is critical, volume-preserving integrators are often used. The second-order Boris algorithm, distinct from RK methods, is the most widely used of these and shows bounded energy error over long integrations \cite{Qin2013, Ripperda2018}. However, because it is second-order, it has accuracy limitations in precise long-time particle-trajectory applications \cite{Qin2013, Winkel2015, Ripperda2018} \add[R1]{which we discuss in Section~\ref{SEC:numerical methods}}. Therefore, we include the fourth-order symplectic Gauss--Legendre Runge--Kutta (RKG) method for comparison, an implicit RK method that preserves the system's Hamiltonian structure over extended integration periods while retaining higher-order trajectory accuracy \cite{Iserles1996, Yugo2001, Calvo2003}.

Higher-order RK methods \cite{Walters2022, Meseguer2025}, variations of the Boris method \cite{Wang2025}, and other improved trajectory-integration schemes \cite{Soni2021, Brizard2022, Huang2022, Umeda2023} remain active areas of research in this field, each implementing different mechanisms or constraints for maintaining accuracy and stability. In this work, we do not attempt to provide a comprehensive survey of numerical algorithms. Instead, we seek to introduce the PS method and demonstrate that it yields low errors in exact invariants over \change[a]{decades}{extended integrations}, with execution times \change[a]{comparable to}{that can be competitive with} those of alternative algorithms. As a result, it can be used to address problems of the type described in the introduction.

The Parker--Sochacki method, a power-series approach for solving initial value problems, builds on the classical Picard method \cite{Parker1996}. In the Picard method, the differential equation is first written in integral form, and the solution is approximated iteratively by substituting the previous approximation into this integral \cite{Boyce2017}. Each iteration refines the approximation from the previous one, giving a power series representation of the exact solution when one exists. The Picard method applies to ODEs of the form $d\mathbf{x}/dt=f(\mathbf{x},t)$ and converges to the exact solution on intervals where the solution possesses a valid power series representation. In special cases, such as Examples I and II in \ref{sect:examples}, the resulting series corresponds to a known closed-form equation. 

%Boyce2017:"One method of showing that the integral equation (3) has a unique solution is known as the method of successive approximations or Picard’s iteration method. In using this method, we start by choosing an initial function φ0, either arbitrarily or to approximate in some way the solution of the initial value problem. The simplest choice is φ0(t) = 0;  then φ0 at least satisfies the initial condition in equations (2), although presumably not the differential equation. The next approximation φ1 is obtained by substituting φ0(s) for φ (s) in the right-hand side of equation (3) and calling the result of this operation φ1(t)."

While the Picard method is theoretically powerful and has high-order accuracy, its practical use has historically been limited by the algebraic complexity of performing successive integrations for nonlinear, nonpolynomial, or coupled ODE systems to obtain a power series representation \cite{Ansari2024, Leal2015, Rudmin1998}. The PS method addresses these limitations by using auxiliary variables and Cauchy product identities to rewrite the nonlinear components of an ODE, thereby reducing algebraic complexity and giving recursive approximations \cite{Parker1996, Guenther2019}. This approach preserves the underlying requirement that the exact solution be analytic for the series to converge, but it makes constructing that power series tractable for a wide range of nonlinear systems. Consequently, the step size is directly linked to the convergence rate of the power-series representation as illustrated in \citeA{Parker1996}, enabling the use of larger time steps. This contrasts with RK methods, where step sizes are typically determined by numerical stability constraints and local error estimates rather than by the analytic properties of the solution.

%Ansari2024: "The Picard method is a recursive algorithm commonly used to solve initial value problems. However, the main challenge of this method is computing the integral of the complex and nonlinear function."
%"this method can become complicated or unsolvable when dealing with nonlinear cases due to the calculation of corresponding integrals."

%Leal2015: "Usually, the application of Picard’s method generates large mathematical expressions difficult to handle."

%Rudmin1998: "Continuing this process constitutes the Picard iteration, which consists of expressing each righthand member of equations (1) through (4) as a power series of order n in t, and then using the equations to increment the number of terms for the series representing each left-hand member. This was published by Picard in 1928 [2], but has been since regarded as an impractical formalism"

%Guenther2019:"The Parker-Sochacki Method (PSM) allows the numerical approximation of solutions to a polynomial initial value ordinary differential equation or system (IVODE) using an algebraic power series method. PSM is equivalent to a modified Picard iteration and provides an efficient, recursive computation of the coefficients of the Taylor polynomial at each step."

%Parker1996: See example 3. 

\add[R1]{The second-order Boris method \cite{Boris1970} is a common method in particle-in-cell simulations where fast integration times and energy conservation are needed and individual-trajectory accuracy is not critical \cite{Birdsall1991}; it has also been used for test-particle studies in radiation-belt fields \cite{Desai2021}. By design, it conserves kinetic energy to round-off levels in a static magnetic field \cite{Qin2013}, though its bounded error requires a fixed time-step \cite{Hairer2006}. We are interested in applications that require accurate individual trajectories, including gyrophase-dependent behavior that results in resonances and determines the transition to chaotic orbits; as discussed in Section 1, our goal is that trajectories remain accurate over many gyroperiods so numerical error is not mistaken for physical effects. This requires that the error in a trajectory be detectable to some extent; we assess this using the relative errors in the exact invariants, kinetic energy and canonical momentum. For the PS and explicit RK methods, neither quantity is forced to be conserved, and their errors are associated with accumulated trajectory error. For the Boris method, they are not, since it is specifically designed to conserve energy. Additionally, although the error in the canonical momentum is not guaranteed to remain bounded in a spatially varying field for the Boris method \cite{Hairer2018}, it did not consistently grow over $10^6$ equatorial gyroperiods in the orbits we examined. Therefore, the Boris method's errors accumulate in the trajectory without any obvious indicators. Moreover, although the Boris step size can be reduced, there is no intrinsic criterion for choosing it. Short of a step-refinement study for each orbit, which is impractical for long integrations over many particles, there is no indication of when the step is small enough. In contrast, in the PS method, machine precision determines the number of terms at each step, and in RK45 the step size is controlled by an embedded local error estimate. The RK4 and RKG methods also use a fixed time step, but an inadequate step size results in growth in the invariant errors. For these reasons, the Boris method is not considered for our long-term integration aims where accurate trajectories are of interest, and is therefore not included in the comparisons that follow.}

We assess the methods using three criteria: accuracy, stability, and computational efficiency. Accuracy is quantified by the relative errors of the exact invariants. Stability refers to the ability of a method to conserve these invariants over extended numbers of equatorial gyroperiods, without showing rapidly increasing relative errors or stalled execution. In Section~\ref{SEC:results}, computational efficiency is evaluated first by comparing runtimes under identical fixed-time-step conditions, and second by comparing the runtime required to meet a minimum target relative kinetic energy error for select runs. The RK4 method is used as a baseline for comparison with a fixed-step PS method. A comparison with the adaptive RK45 method is used to assess the performance of the fixed-step PS approach relative to an adaptive-step solver. The fixed-step RKG is included as a fourth-order symplectic reference.

\section{Implementation of ODE Solvers}
\label{sect:implementation}
We implement \change[R2]{the fixed-step RK and PS}{all} methods in Python using 64-bit floating-point numbers \cite{IEEE754}, NumPy \cite{harris2020} arrays, and Numba \cite{lam2015} to accelerate performance-critical loops. For the PS method, each dynamical and auxiliary variable is represented as a truncated power series in a dimensionless time variable. The PS recurrence relations, explicitly derived in Section~\ref{sec:dipole} for the dipole problem, are used to compute the power series coefficients recursively up to a maximum order of 40. In practice, an adaptive truncation criterion terminates the summation at a smaller order based on the condition that if the ratio of the next term in the sum to the sum of all previous terms is less than machine epsilon, the summation is terminated.

In Section~\ref{sec:fixed time step}, we compare the PS method to a fixed-step RK4 method, the adaptive-step RK45 method \cite{Dormand1980}\change[R2]{as implemented by SciPy \cite{scipy}}{, which we implemented in Numba following the SciPy implementation of the Dormand--Prince method \cite{scipy}}, and the two-stage, fourth-order, symplectic RKG method (identified as RKG2 in \citeA{Yugo2007}). \remove[R2]{The SciPY RK45 algorithm is included as a widely accessible, off-the-shelf capability for comparison. Because it is not implemented in Numba, its execution times are slower. While an optimized implementation would achieve runtimes comparable to the PS and RKG methods, the accuracy would remain unchanged.} Following the approach of \citeA{Yugo2007}, we solve the stage equations by a fixed-point iteration \cite{Calvo2003}. The iteration tolerances and limits are not reported in \citeA{Yugo2001, Yugo2007}; we therefore chose values consistent with the reported step size ($65$ steps per gyroperiod) and compared our RKG relative kinetic energy error results with theirs (canonical momentum values were not compared because they were not reported). We iterate until the largest change in the stage values between successive sweeps falls below $10^{-15}$, capping at $100$ sweeps, though roughly 10 sweeps were typically required. This approach resulted in error levels approximate to those reported in \cite{Yugo2007} for the time interval. All fixed-step solvers (RK4, RKG, and PS) are run with the same time step, ensuring that differences in runtime and invariant conservation arise from the methods themselves rather than from step-size adaptivity; RK45, which selects its own step size, is included as an adaptive-step reference. For RK45, the relative and absolute tolerances are \change[R2]{$10^{-8}$ and $10^{-10}$}{$10^{-18}$ and $10^{-20}$}, respectively\add[R2]{, to optimize RK45 accuracy}. \add[R2]{Because the embedded error estimate is formed from a weighted sum of the stage derivatives, tolerances below machine precision continue to reduce the step size and the accumulated error. In a tolerance sweep over the particles considered here, we found that $10^{-18}$ was the smallest value at which the step-size controller operated reliably in all cases; smaller values produced frequent step rejections without consistently reducing the invariant errors. However, these strict tolerances result in substantially longer runtimes.} For the PS method, the power-series expansions are derived analytically; beyond rewriting the equations of motion in dimensionless form, no further time or field rescaling is applied. All trajectory calculations use Cartesian coordinates.

%Note that Yugo and Iyemori aren't really explicit in how they solved for the RKG; however they cite Calvo which uses the fixed-point iteration. Comparing code vs papers vs results, we repeated the best we could with what information we had and I feel we are upfront about that. Our results are fairly consistent with theirs.

We report wall-clock computation times to indicate the relative scaling behavior of the integrators rather than to establish a definitive performance ranking; absolute times are hardware- and software-dependent and reflect our implementations rather than the computational limits of any method.
%All integrators allow for further optimizations that we did not explore, such as alternative starting algorithms for the RKG iteration or step-size tuning for the PS method.

\remove[a]{Generative AI tools were used for late-stage revisions of text and code. Claude was used to (1) modify the style and annotations of existing plotting code and (2) refactor and reorganize existing code without modifying its functionality or algorithms; the refactored code is the version archived in the repository described in the Open Research Section, and all results were regenerated with it. Grammarly was used for minor revisions of grammar, spelling, punctuation, and diction in text written and revised by the authors. The authors take full responsibility for the content of the manuscript and the archived code.}\note[a]{This was moved to the end of the document}

\section{PS Method for the Dipole Problem}\label{sec:dipole}

In this section, we derive the power series relations needed to solve the dipole problem. In ~\ref{sect:examples}, we show the derivation of power series for simpler problems to provide additional background on the approach.

The dynamics of a relativistic charged particle in a static magnetic dipole field are governed by the Lorentz force law,
        \begin{equation*}
           m_{\text{rel}}\frac{d\mathbf{v}}{dt}=q\,\mathbf{v}\times\mathbf{B},
        \end{equation*}
where $q$ is the particle's charge, $\mathbf{B}$ is the magnetic field, and $m_{\text{rel}}$ is the relativistic mass. The particle's position evolves according to $d\mathbf{r}/dt=\mathbf{v}$, where $\mathbf{r}$ and $\mathbf{v}$ are the three-dimensional position and velocity vectors of the particle. The relativistic mass, $m_{\text{rel}} = m_o/\sqrt{1-v^2/c^2}$, is treated as a constant because the rest mass, $m_o$, is a constant by definition and $v^2$ is a constant \cite{Dragt1965,Yugo2007}.
%Since the magnetic field does no work on the particle, the scalar speed remains constant, allowing the relativistic mass to be treated as a constant.
%Since $\gamma$ is constant in this formulation, conservation of $E$ is equivalent to conservation of the speed, and hence of the relativistic kinetic energy. For a small drift in speed, the corresponding relative drift in the relativistic kinetic energy exceeds that of $E$ by a factor of $\gamma(\gamma+1)/2$; because this factor depends only on the particle and not on the integrator, it is common to all methods for a given run and does not affect any comparison between them.

To write the equations in dimensionless form, we define dimensionless position and velocity vectors by introducing a characteristic length, $r_0$, and speed,
$v_0$, such that
        \begin{equation*}
            \mathbf{r'}=\frac{\mathbf{r}}{r_0}\quad\mbox{and}\quad
            \mathbf{v'}=\frac{\mathbf{v}}{v_0}.
        \end{equation*}
We take $r_0 = R_E$, one Earth radius, and normalize the magnetic field by the equatorial field strength at that distance,
        \begin{equation*}
            B_0 = \frac{\mu_0 \mathfrak{m}}{4\pi R_E^3},
        \end{equation*}
where $\mu_0$ is the vacuum permeability and $\mathfrak{m}$ is the magnitude of the dipole moment. Time is normalized by the inverse of the angular gyrofrequency associated with $B_0$,
        \begin{equation*}
            \tau\equiv\frac{t}{\tau_0},\qquad\text{where}\quad
            \tau_0\equiv\frac{m_{\text{rel}}}{|q|B_0}\;,
        \end{equation*}
and $v_0$ chosen such that $r_0/v_0=\tau_0$. The dimensionless equations of motion are then
        \begin{align}
            \frac{d\mathbf{v'}}{d\tau} &= s\,\mathbf{v'}\times \mathbf{B'}
            \label{EQN: Rep Lorentz}\\
        \intertext{and}
            \frac{d\mathbf{r'}}{d\tau} &= \mathbf{v'},
            \label{EQN: Rep dr/dt}
        \end{align}
where $s \equiv q/|q|$ is the sign of the charge, and the dimensionless dipole field is
        \begin{equation}
            \mathbf{B'}(\mathbf{r'})=\frac{\mathbf{B}}{B_0}
            =\frac{3(\hat{\boldsymbol{\mathfrak{m}}}\cdot\hat{\mathbf{r}})\hat{\mathbf{r}}
            - \hat{\boldsymbol{\mathfrak{m}}}}{r'^3},
            \label{EQN: Dipole B main}
        \end{equation}
where $\hat{\boldsymbol{\mathfrak{m}}}$ is the unit vector along the dipole axis. We align the dipole axis with the $-z$-axis, $\hat{\boldsymbol{\mathfrak{m}}} = -\hat{\mathbf{z}}$, corresponding to a southward-pointing Earth-like oriented dipole. In Cartesian coordinates Eq.~\eqref{EQN: Dipole B main} becomes
        \begin{equation}
            B'_x = -\frac{3x'z'}{r'^5}, \qquad
            B'_y = -\frac{3y'z'}{r'^5}, \qquad
            B'_z = -\frac{2z'^2 - x'^2 - y'^2}{r'^5},
            \label{EQN: Dipole B components}
        \end{equation}
where $r' = \sqrt{x'^2 + y'^2 + z'^2}$. 

The fixed time step is referenced to the gyroperiod at the local magnetic equator, where the field is weakest along a given field line and the gyroperiod is longest. For the field line crossing the magnetic equator at radial distance $L$ (the magnetic shell parameter in units of $R_E$), the equatorial field magnitude is $B_{\mathrm{eq}} = B_0/L^3$. Therefore, the dimensionless equatorial gyroperiod is
        \begin{equation*}
            T = \frac{2\pi B_0}{B_{\mathrm{eq}}} = 2\pi L^3.
        \end{equation*}
The resulting dimensionless step size is
        \begin{equation}
            \Delta \tau = \frac{2\pi L^3}{N},
            \label{eqn: dipole tau}
        \end{equation}
where $N$ is the number of integration steps per equatorial gyroperiod. Because the local gyroperiod scales inversely with the field magnitude, the off-equatorial portions of a trajectory are resolved by proportionally fewer steps per gyration. In the guiding center approximation, the shortest local gyroperiod occurs at the mirror points, where $B = B_{\mathrm{eq}}/\sin^2\alpha_{\mathrm{eq}}$ for equatorial pitch angle $\alpha_{\mathrm{eq}}$, so for the smallest pitch angle considered here ($30^\circ$) every gyration is resolved by no fewer than $N/4$ steps.

\subsection{Recurrence Relations}
From this point forward, all variables are understood to be dimensionless, and the prime notation is omitted. We begin by expressing the $x$-component of position and velocity as power series in time:
        \begin{equation}\label{EQN: ConstB x}
            x=\sum_{i=0}^\infty x_i\tau^i
        \end{equation}
        \begin{equation}\label{EQN: ConstB v}
            v_x=\sum_{i=0}^\infty v_{xi}\tau^i .
        \end{equation}
Substituting Eqs.~\eqref{EQN: ConstB x} and \eqref{EQN: ConstB v} into the
$x$-component of Eq.~\eqref{EQN: Rep dr/dt} gives
    \begin{align*}
            \frac{d}{d\tau}\left(\sum_{i=0}^\infty x_i\tau^i\right)=& \sum_{i=0}^\infty v_{xi}\tau^i\;,\\
\intertext{which reduces to}
            \sum_{i=0}^\infty ix_i\tau^{i-1}=&\sum_{i=0}^\infty v_{xi}\tau^i.
    \end{align*}
We reindex the left-hand side using $i \to i+1$ and equate powers of $\tau$ giving
    \begin{equation*}
        x_{i+1}=\frac{v_{xi}}{i+1}.
    \end{equation*}
This generates the coefficients for Eq.~\eqref{EQN: ConstB x}. Similarly, the remaining position recurrence relations are
        \begin{align*}
            y_{i+1}=\frac{v_{yi}}{i+1}\quad
            \mbox{and}\quad
            z_{i+1}=\frac{v_{zi}}{i+1}.
        \end{align*}
The remaining recurrence relations rely on two power series identities. Products of series are combined using the Cauchy product identity \cite{Knuth1997},
        \begin{equation}
            \left(\sum_{i=0}^\infty P_i\tau^i\right)\left(\sum_{i=0}^\infty Q_i\tau^i\right)
            =\sum_{i=0}^\infty\tau^i\sum_{j=0}^i P_jQ_{i-j}.
            \label{EQN:cauchyb}
        \end{equation}
For the ratio of two power series, $W=P/Q$, rearranging as $QW=P$ and applying Eq.~\eqref{EQN:cauchyb} to the left-hand side gives
    \begin{align*}
        \sum_{j=0}^i Q_jW_{i-j}=&\,P_i.\\
\intertext{Isolating the $j=0$ term from the summation gives}
        Q_0W_i+\sum_{j=1}^i Q_jW_{i-j}=&\,P_i,
    \end{align*}
and rearranging results in the identity
        \begin{equation}
            W_i=\frac{1}{Q_0}\left(P_i-\sum_{j=1}^i Q_jW_{i-j}\right).
            \label{EQN: Cauchy a/b}
        \end{equation}
We define a set of auxiliary variables ($a,\,b,\,c,\,d,\,e,\,f,\,g,$ and $u$) by first letting
        \begin{align*}
            a &\equiv \frac{1}{r^5}, \\
            b &\equiv 2z^2-x^2-y^2, \\
            c &\equiv yz,\\
         \intertext{and}
            d &\equiv xz,
        \end{align*}
such that
        \begin{align*}
           B_x&=-\frac{1}{r^5}3xz=-3ad,\\
           B_y&=-\frac{1}{r^5}3yz=-3ac,\\
         \intertext{and}
            B_z&=-\frac{1}{r^5}(2z^2-x^2-y^2)=-ab.
        \end{align*}
Next, we express the acceleration equations in terms of these auxiliary variables and define the remaining auxiliary variables,
        \begin{align*}
            \frac{dv_x}{d\tau} &= s\left(B_z v_y - B_y v_z\right)
            = sa(3c v_z - b v_y) = sae, \\
            \frac{dv_y}{d\tau} &= s\left(B_x v_z - B_z v_x\right)
            = sa(b v_x - 3d v_z) = saf, \\
         \intertext{and}
            \frac{dv_z}{d\tau} &= s\left(B_y v_x - B_x v_y\right)
            = sa(3d v_y - 3c v_x) = sag,
        \end{align*}
where
        \begin{align*}
            e &\equiv 3c v_z - b v_y, \\
            f &\equiv b v_x - 3d v_z, \\
         \intertext{and}
            g &\equiv 3d v_y - 3c v_x.
        \end{align*}
Let $u\equiv r^2$ such that
        \begin{gather*}
            	    a=u^{-\frac{5}{2}},\\
            \intertext{and using}
            	\ln a=-\frac{5}{2}\ln u\\
             \intertext{gives}
            	\frac{1}{a}\frac{da}{d\tau}=-\frac{5}{2}\frac{1}{u}\frac{du}{d\tau}\\
            \intertext{which can be rearranged into an expression containing $a/u$, a series divided by another series,}
            		\frac{da}{d\tau}=-\frac{5}{2}\frac{a}{u}\frac{du}{d\tau}.\\
            \intertext{Letting $\zeta=a/u$, substituting in the summations for each auxiliary variable, and reindexing the derivative terms gives }
            		\sum_{i=0}^\infty (i+1)a_{i+1}\tau^i=-\frac{5}{2}\left( \sum_{i=0}^\infty\zeta_i\tau^i \right)\left( \sum_{i=0}^\infty (i+1)u_{i+1}\tau^i \right), \\
            \intertext{where the terms on the right can be simplified using the Cauchy product identity, Eq.~\eqref{EQN:cauchyb},}
                    	\sum_{i=0}^\infty (i+1)a_{i+1}\tau^i=-\frac{5}{2}\sum_{i=0}^\infty \tau^i\left(\sum_{j=0}^i(i-j+1)\zeta_j u_{i-j+1}\right).\\
            \intertext{Matching powers of $\tau$ gives}
                    a_{i+1}=-\frac{5}{2(i+1)}\sum_{j=0}^i(i-j+1)\zeta_j u_{i-j+1},\\
            \intertext{and from Eq.~\eqref{EQN: Cauchy a/b}, with $P=a$ and $Q=u$, we have}
                \zeta_i=\frac{1}{u_0}\left(a_i-\sum_{j=1}^iu_j\zeta_{i-j}\right)\quad\mbox{where}\quad \zeta_0=\frac{a_0}{u_0}.
        \end{gather*}
Applying the same process, the recurrence relations for the remaining auxiliary variables are
        \begin{align*}
            u_i=&\sum_{j=0}^i x_jx_{i-j}+\sum_{j=0}^i y_jy_{i-j}+\sum_{j=0}^i z_jz_{i-j},\\
            b_i=&\,2\sum_{j=0}^i z_jz_{i-j}-\sum_{j=0}^i x_jx_{i-j}-\sum_{j=0}^i y_jy_{i-j},\\
            c_i=&\sum_{j=0}^i y_jz_{i-j},\\
            d_i=&\sum_{j=0}^i x_jz_{i-j},\\
            e_i=&\,3\sum_{j=0}^i c_jv_{z,i-j}-\sum_{j=0}^i b_jv_{y,i-j},\\
            f_i=&\sum_{j=0}^i b_jv_{x,i-j}-3\sum_{j=0}^i d_jv_{z,i-j},\\
    \intertext{and}
            g_i=&\,3\sum_{j=0}^i d_jv_{y,i-j}-3\sum_{j=0}^i c_jv_{x,i-j}.
        \end{align*}
The velocity recurrence relations can then be expressed as
        \begin{align*}
            v_{x, i+1} &=  \frac{s}{i+1} \sum_{j=0}^i a_j e_{i-j}, \\
            v_{y, i+1} &=   \frac{s}{i+1} \sum_{j=0}^i a_j f_{i-j}, \\
      \intertext{and}
            v_{z, i+1} &=   \frac{s}{i+1} \sum_{j=0}^i a_j g_{i-j}.
        \end{align*}
At each time step, the zeroth-order coefficients ($x_0$, $v_{x0}$, $u_0$, and so on) are evaluated from the state at the beginning of the step, and the recurrence relations above then generate all higher-order coefficients. The updated state follows from evaluating the truncated series at $\tau = \Delta\tau$. At the end of each step, the auxiliary variables are re-evaluated, or tethered, directly from the advanced position and velocity rather than advanced by their own summed series, keeping them consistent with the dynamical state at the start of the next step (see Appendix~\ref{Intro: ExIII} for details).

\section{Results}\label{SEC:results}

\subsection{Comparisons at a Common Fixed Time Step}\label{sec:fixed time step}
For the following simulations, $N=65$ is used for each fixed-step method, corresponding to 65 integration steps per gyroperiod at the magnetic equator, where the gyroperiod is longest. Particles are initialized at $5\,R_E$, giving a fixed time step, from Eq.~\eqref{eqn: dipole tau}, of $\Delta\tau=2\pi(5^3)/65\approx12.1$. Simulation durations are expressed in terms of the number of elapsed equatorial gyroperiods, $\tau/T$. \remove[R2]{Initially, simulations were run for $\tau/T=10^{6}$; however, both RK4 and RK45 frequently became unstable due to high kinetic energy errors. Additionally, RK45 underwent repeated step-size refinements that led to excessive computational runtime and, in several instances, stalling execution.} To allow for meaningful comparison and to capture the long-time trends of the RK methods, each simulation was run for a common time interval of $\tau/T = 1.5\times 10^{5}$. \remove[R2] {This allowed most RK4 and RK45 runs to complete before instability (see Tables~\ref{tab:proton_energy_results} and \ref{tab:electron_energy_results} in \ref{appendix:tables}).} Additional long-time analyses using extended gyroperiod counts were performed for the PS and RKG methods and are discussed in the corresponding subsections. 

Accuracy is defined by the relative errors of the two exact invariants, kinetic energy, $E$, and canonical momentum, $P_{\phi}$. The relative kinetic energy error is
\begin{equation*}
    \frac{|\Delta E|}{E_0}=\frac{|E-E_0|}{E_0}
    \quad\mbox{where}\quad
    E = \tfrac{1}{2}\left(v_x^2+v_y^2+v_z^2\right),
\end{equation*}
\noindent and the relative canonical momentum error is
\begin{equation*}
    \frac{|\Delta P_{\phi}|}{|P_{\phi,0}|}=\frac{|P_\phi-P_{\phi,0}|}{|P_{\phi,0}|}
    \quad\mbox{where}\quad
    P_\phi = \rho\, v_\phi - s\,\frac{\rho^2}{r^3},
\end{equation*}
$\rho = \sqrt{x^2 + y^2}$ is the cylindrical radius, $r = \sqrt{x^2 + y^2 + z^2}$, $v_\phi = (x v_y - y v_x)/\rho$ is the azimuthal velocity, $s = q/|q|$ is the sign of the charge, and $P_{\phi,0}$ is the initial canonical momentum. 
%Since $\gamma$ is constant in this formulation, conservation of $E$ is equivalent to conservation of the speed, and hence of the relativistic kinetic energy. For a small drift in speed, the corresponding relative drift in the relativistic kinetic energy exceeds that of $E$ by a factor of $\gamma(\gamma+1)/2$; because this factor depends only on the particle and not on the integrator, it is common to all methods for a given run and does not affect any comparison between them.

For the dipole field, the gradient scale length at the equatorial crossing of the field line is $L_B = L/3$, and the dimensionless equatorial gyroradius is $r_g = v_{\perp} L^{3}$, where $v_{\perp}$ is the dimensionless perpendicular speed at the equator. The parameter $\varepsilon$ introduced in Section~\ref{SEC:intro}, evaluated at the equator, is therefore

\begin{equation*}
    \varepsilon = \frac{r_g}{L_B} = 3\, v_{\perp} L^{2}.
\end{equation*}

In the simulations discussed below, $\varepsilon$ ranges from ${\sim}3\times10^{-3}$ to ${\sim}0.2$; the highest-energy runs of both species therefore lie in a regime where the gyroradius is no longer small compared with the length scale over which the field varies, and the validity of the guiding center approximation, while not used here, is unknown.

For the PS method, the mean truncation order is used to label each run; for example, a simulation truncated on average at the 13th order is denoted PS13.

\subsubsection{Proton Simulations}\label{Proton Simulations}
This section details the simulation of a proton in a magnetic dipole field at four energies, 10~keV; 100~keV; 1~MeV; and 10~MeV and two pitch angles, 90$^{\circ}$ and 30$^{\circ}$. The 90$^{\circ}$ pitch angle isolates the azimuthal drift, while the 30$^{\circ}$ pitch angle allows for both bounce and azimuthal drift motion.

\begin{figure}[!htbp]
    \centering
    \includegraphics[width=0.60\textwidth]{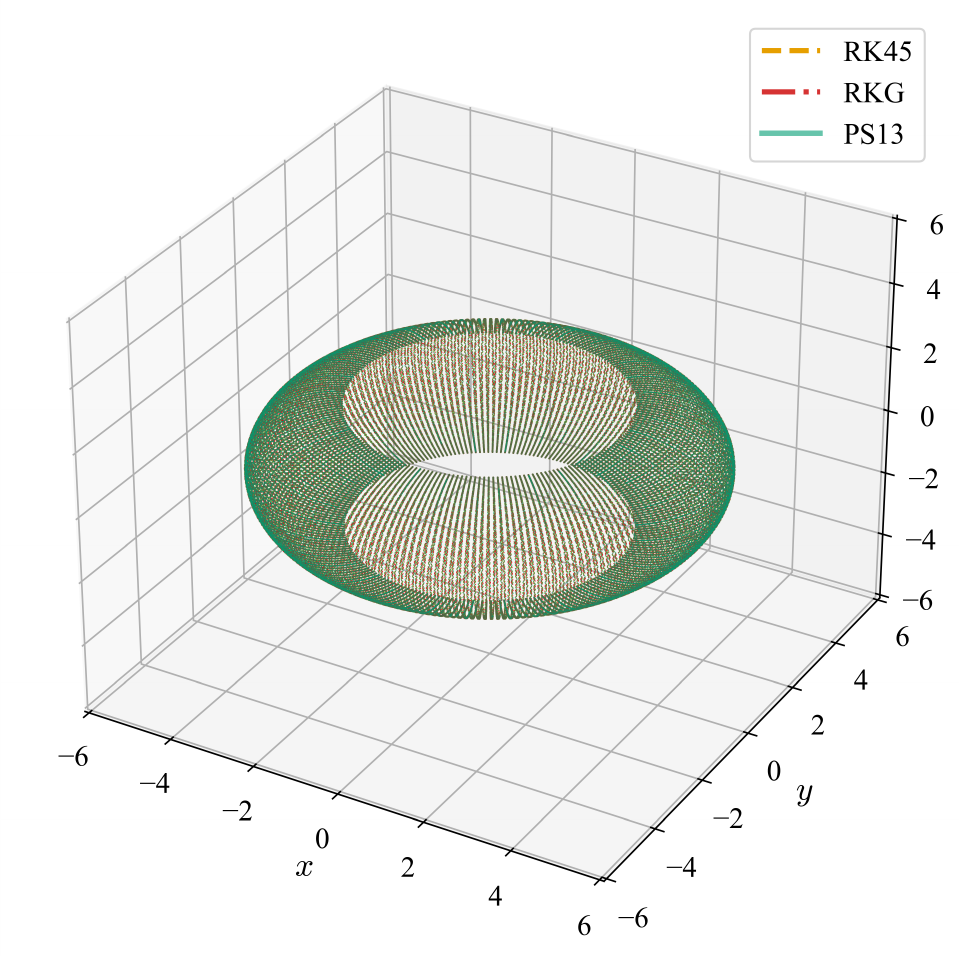}
    \caption{Trajectory of a 100~keV proton in Earth's magnetic dipole field over the final drift period (103.4~minutes) of the common time interval, $\tau/T=1.5\times10^{5}$ (657.1~minutes). Using PS13, RK45, and RKG, the proton is initialized at $5\,R_E$ with a $30^{\circ}$ pitch angle and integrated with a fixed time step corresponding to 65 integration steps per equatorial gyroperiod ($\Delta\tau = 12.1$). RK4 failed due to high relative kinetic energy error at approximately $\tau/T=6\times 10^{3}$ and is therefore omitted. \note[R2]{Image Regenerated}}
    \label{FIG:DipoleB_ex1_traj}
\end{figure}

Figure~\ref{FIG:DipoleB_ex1_traj} shows representative trajectories for PS13, RK45, and RKG for a 100~keV proton with a $30^{\circ}$ pitch angle, over the final drift period (103.4~minutes of physical time) of the common time interval (657.1~minutes of physical time). RK4 is omitted because it failed at approximately $\tau/T=6\times 10^{3}$, with relative kinetic energy errors on the order of $10^6$. By the end of the common time interval,\change[R2]{RK45 has a visible difference in azimuthal position relative to RKG and PS13, which}{ all remaining methods} are in visual agreement.

\begin{figure}[!htbp]
    \centering
    \includegraphics[width=\textwidth]{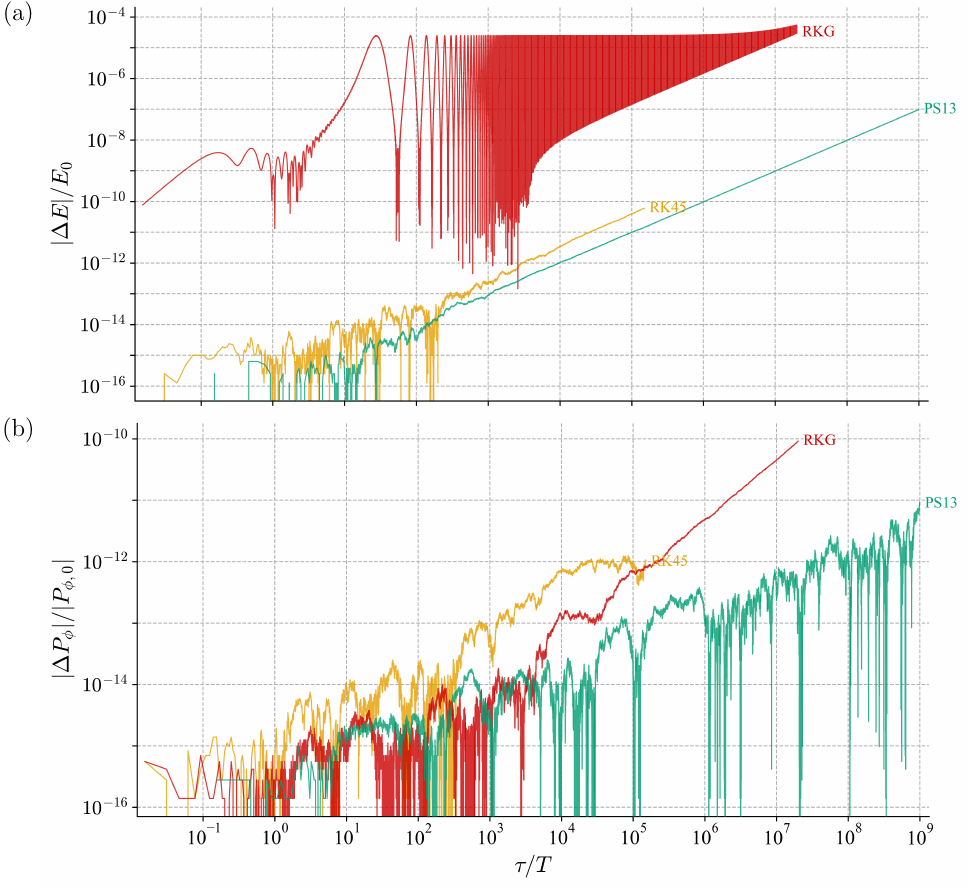}
    \caption{Relative errors of the exact invariants for the proton in Fig.~\ref{FIG:DipoleB_ex1_traj}: (a) kinetic energy and (b) canonical momentum. The RKG and PS13 runs are extended beyond the common interval to $\tau/T=2\times 10^{7}$ (60.8~days) and $\tau/T=10^{9}$ (8.3~years), respectively. \note[R2]{Image Regenerated}} 
    \label{FIG:Dipole_proton_all}    
\end{figure}

The corresponding relative errors of the exact invariants are shown in Fig.~\ref{FIG:Dipole_proton_all}: kinetic energy in panel~(a) and canonical momentum in panel~(b). The figure includes both the common time interval and extended integrations for the RKG and PS13 runs. Over the common time interval, PS13 has relative kinetic energy errors that are \change[R2]{approximately 10 orders of magnitude lower than those}{within an order of magnitude} of RK45 and approximately 5 orders of magnitude lower than those of RKG. Similarly, PS13 has relative canonical momentum errors that are \remove[R2]{more than 7 orders of magnitude lower than those of RK45 and} approximately an order of magnitude lower than those of RKG \add[R2]{and RK45}. \remove[a]{At the common fixed time step, PS13 runtimes are comparable to those of RKG, while RK45 has substantially longer runtimes.} All relative errors and runtimes over the common time interval are summarized in Appendix Table~\ref{tab:proton_energy_results}.

Beyond the common time interval, the extended integrations reveal distinct long-time behaviors. The RKG run, extended to $\tau/T=2\times 10^{7}$ (60.8~days of physical time), shows a gradual secular increase in relative kinetic energy error after $\tau/T=10^{6}$, indicating a degradation of its symplectic behavior. The PS13 run, extended to $\tau/T=10^{9}$ (8.3~years of physical time), shows the relative errors of both exact invariants continuing to grow approximately linearly with simulated time, with no abrupt loss of accuracy or stability. If this trend were to hold for an additional order of magnitude, corresponding to approximately 80 years of physical time, the PS13 error would remain below $10^{-6}$. RKG was not extended further due to resource constraints\add[R2]{, because of its long runtimes at the tolerance values used, RK45 was not extended beyond the common time interval}.

\begin{figure}[!htbp]
    \centering
    \includegraphics[width=0.90\textwidth]{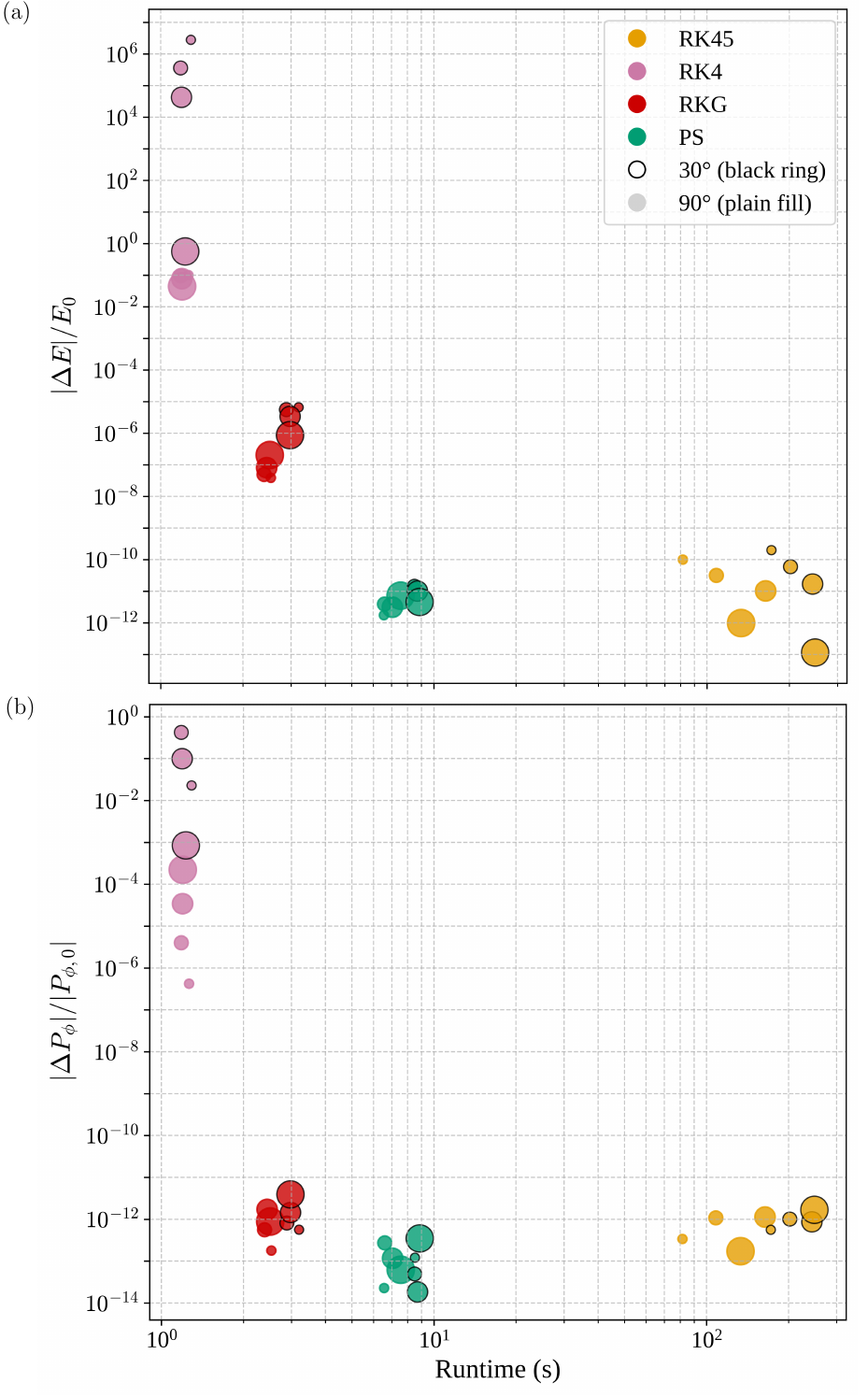}
    \caption{Comparison of the PS, RK4, RK45, and RKG methods for protons in Earth's magnetic dipole field, initialized at $5\,R_E$. The fixed-step methods (PS, RK4, and RKG) use an identical time step of 65 integration steps per equatorial gyroperiod ($\Delta\tau = 12.1$), while RK45 uses adaptive stepping. All results are evaluated over the common time interval of $\tau/T = 1.5\times 10^{5}$. Marker size indicates proton energy (10~keV, 100~keV, 1~MeV, and 10~MeV), with larger markers corresponding to higher energies; black-ringed markers denote the $30^{\circ}$ pitch angle runs and plain filled markers the $90^{\circ}$ runs. Each point is the mean over the final 1\% of the run duration of (a) the relative kinetic energy error, $|\Delta E|/E_0$, and (b) the relative canonical momentum error, $|\Delta P_{\phi}|/|P_{\phi,0}|$, plotted against computational runtime. The legend in panel~(a) applies to both panels.\note[R2]{Image Regenerated}}
    \label{FIG:bubbleplot_protons}
\end{figure}

Figure~\ref{FIG:bubbleplot_protons} shows the relationship between invariant accuracy and computational runtime in our configuration: kinetic energy in panel~(a) and canonical momentum in panel~(b). Each marker represents the mean relative error computed over the final 1\% of the run.  \remove[R2]{Markers absent from Fig.~\ref{FIG:bubbleplot_protons} correspond to runs that did not complete the common time interval due to numerical instability.} A complete tabulation of these metrics is provided in Table~\ref{tab:proton_energy_results} of \ref{appendix:tables}. All methods except RK45 use the common fixed time step, isolating differences in how numerical error accumulates within each method rather than differences arising from step-size selection. For PS runs, the truncation criterion from Section~\ref{sect:implementation} was used; for all runs, this ranged between 11 and 13; see \ref{appendix:tables} for details.

RK4 has the shortest runtimes at the expense of the largest errors in both invariants, producing large kinetic energy errors for pitch angles of 90$^\circ$ and failing due to instabilities for most protons at 30$^\circ$. RK45 \change[R2]{improves on RK4's accuracy but still conserves neither invariant well, and does so at the}{has accuracy competitive with PS, but with the} longest runtimes of any method\add[R2]{ due to its tolerance levels. To obtain a comparable runtime, RK45 tolerance levels could be set at approximately $10^{-11}$ but its relative errors increase by several orders of magnitude}. RKG occupies an intermediate regime with runtimes comparable to the RK4 method; it conserves the canonical momentum to near machine precision, within about an order of magnitude of the PS method's; however, its relative kinetic energy error\change[R2]{, though far lower than that of the RK methods,}{} remains \change[R2]{roughly six}{several} orders of magnitude above that of the PS method and grows secularly on long timescales. By contrast, the PS method \change[R2]{is the only method to have the lowest errors in both exact invariants simultaneously, at runtimes comparable to RKG's and far below those of RK45}{has low errors in both exact invariants in every case, with runtimes far below those of RK45}.

In Section~\ref{SEC:Dip Discussion} that follows, a comparison using a minimum target relative kinetic energy error rather than matched step size further reinforces these results. When the parameters of \change[a]{RK4 and PS}{the methods} are chosen to have comparable relative kinetic energy errors, the PS method reaches the desired accuracy requirements with runtimes that are \change[R2]{approximately an order of magnitude faster than RK4 in our system configuration}{several times faster than the explicit methods and comparable to RKG}.

\newpage

\clearpage

\clearpage

\subsubsection{Electron Simulations}
This section details the simulation of an electron in a magnetic dipole field at four energies: 50~keV, 1~MeV, 100~MeV, and 150~MeV, and two pitch angles: 90$^{\circ}$ and 60$^{\circ}$, the latter chosen to match the electron configuration of \citeA{Yugo2007}. As in the proton simulations, the 90$^{\circ}$ pitch angle isolates the azimuthal drift, while the 60$^{\circ}$ pitch angle has bounce motion along the field line and azimuthal drift. \remove[a]{The RKG method was unstable for every electron configuration considered here; its relative kinetic energy error exceeded unity within the first few time steps, consistent with the findings of \mbox{\citeA{Yugo2007}} for electrons in Cartesian coordinates. It is therefore omitted from the electron comparisons.}\add[a]{The RKG method remained stable for every electron configuration considered here, though it continued to exhibit drift over longer trajectories, and is therefore included in the electron comparisons. This differs from \mbox{\citeA{Yugo2007}}, who reported failures for electrons in Cartesian coordinates for their step size selections.}

%Yugo2007:"..relativistic electrons having energies from 10 to 150 MeV with pitch angles from 90° to 30° at L ∼ 5 are also practiced. For the electrons above 50 MeV, the calculations are successful for 10,000,000 steps with the time step 0.1 × R_E/c by fourth and sixth RKG with cylindrical coordinates. The calculations by standard Runge-Kutta method are failed in any coordinate system, and the calculations by fourth or sixth RKG with Cartesian or polar coordinates.."

\begin{figure}[!htbp]
    \centering
    \includegraphics[width=.6\textwidth]{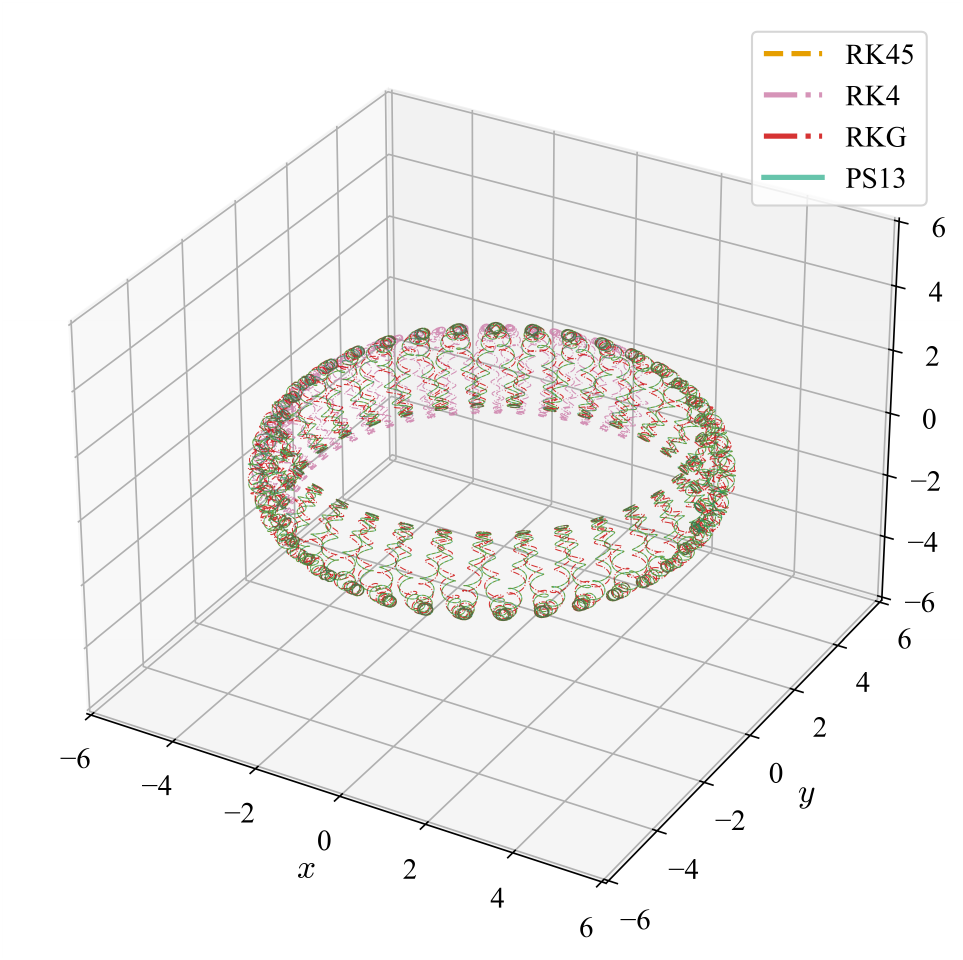}
    \caption{Same as Figure~\ref{FIG:DipoleB_ex1_traj} but for a 100~MeV electron with an initial pitch angle of $60^{\circ}$, shown over the final drift period (11.6~s) of the common time interval, $\tau/T=1.5\times10^{5}$ (70.4~minutes). \remove[a]{RKG is omitted due to its instability for electrons in Cartesian coordinates.}\note[R2]{Image regenerated}}
    \label{FIG:DipoleB_ex2_traj}
\end{figure}

Figure~\ref{FIG:DipoleB_ex2_traj} shows representative trajectories for PS13, RK4,\add[a]{ RKG,} and RK45 for a 100~MeV electron with a $60^{\circ}$ pitch angle, evaluated over the final drift period (11.6~s) of the common time interval (70.4~minutes). By the end of the common time interval, \change[R2]{both RK4 and RK45 have visible differences in azimuthal position relative to PS13}{RK45 and PS13 remain in visual agreement while RKG and RK4 have noticeable trajectory drifts}.

\begin{figure}[!htbp]
    \centering
    \includegraphics[width=\textwidth]{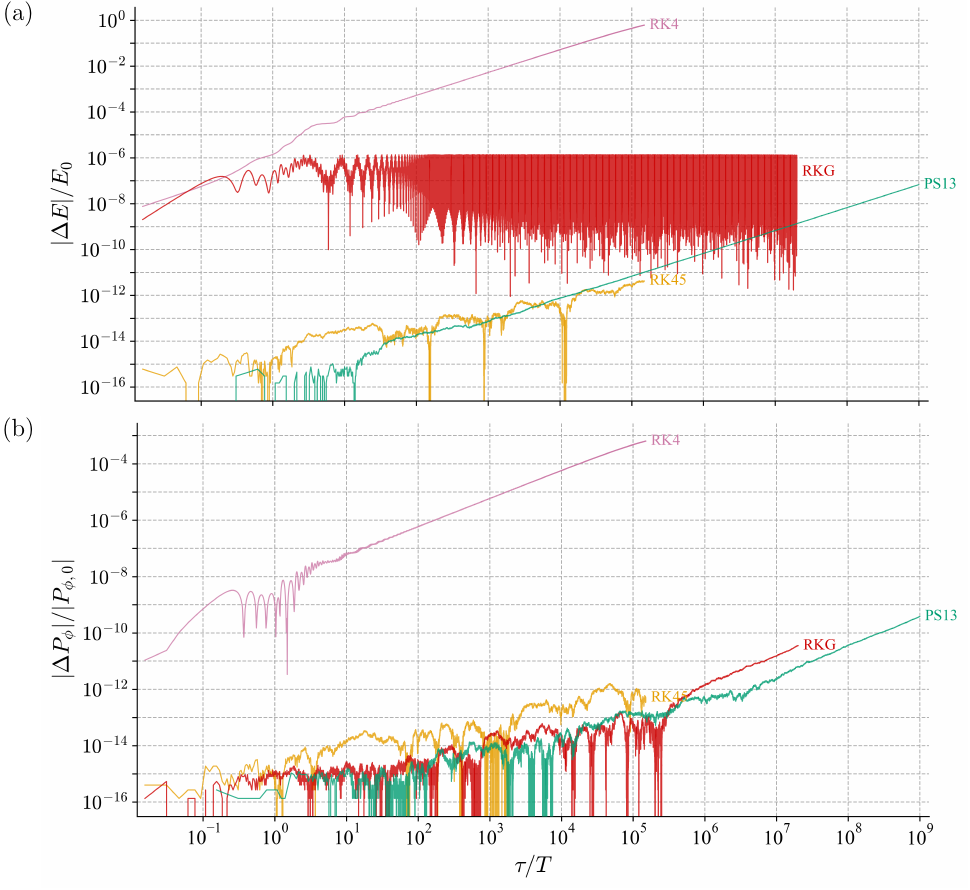}
    \caption{Relative errors of the exact invariants for the electron of Fig.~\ref{FIG:DipoleB_ex2_traj}: (a) kinetic energy and (b) canonical momentum. \change[a]{The PS13 run is extended beyond the common interval to $\tau/T=10^9$ ($\sim$325.8~days).}{The RKG and PS13 runs are extended beyond the common interval to $\tau/T=2\times10^{7}$ (6.5~days) and $\tau/T=10^{9}$ (325.8~days), respectively.}\note[R2]{Image regenerated}} \label{FIG:Dipole_electron_all}
\end{figure}

The corresponding relative errors of the exact invariants are shown in Fig.~\ref{FIG:Dipole_electron_all}: kinetic energy in panel~(a) and canonical momentum in panel~(b). The figure includes both the common time interval and the extended PS13 \add[a]{and RKG} integration\add[a]{s}. Over the common time interval, \change[R2]{all methods have relative kinetic energy errors that increase approximately linearly in time; however, PS13 has relative kinetic energy errors that are approximately 10 orders of magnitude lower than those of RK4 and RK45, and relative canonical momentum errors that are 8 to 9 orders of magnitude lower. At the common fixed time step, PS13 runtimes are longer than those of RK4 but substantially shorter than those of RK45.}{. PS13 has relative kinetic energy errors that are approximately 11 orders of magnitude lower than those of RK4 and up to 7 orders of magnitude lower than those of RKG, and comparable to those of RK45. Relative canonical momentum errors are approximately 10 orders of magnitude lower than those of RK4 and comparable to those of RK45 and RKG. At the common fixed time step, PS13 runtimes are longer than those of RK4 and RKG but substantially shorter than those of RK45.} All relative errors and runtimes over the common time interval are summarized in Appendix Table~\ref{tab:electron_energy_results}.

Beyond the common time interval, the PS13 run, extended to $\tau/T=10^{9}$ (325.8~days of physical time), shows the relative errors of both exact invariants continue to grow approximately linearly with simulated time, with no abrupt loss of accuracy or stability. \add[a]{The RKG run, extended to $\tau/T=2\times10^{7}$ (6.5~days of physical time), maintains a bounded relative kinetic energy error, while its relative canonical momentum error grows approximately linearly after $\tau/T=10^{6}$. RKG was not extended further due to resource constraints,}\add[R2]{ and because of its long runtimes at the tolerances used, RK45 was not extended beyond the common time interval.} As in the proton simulations, these results demonstrate that the PS method maintains long-time accuracy and stability even for highly relativistic electrons.

\begin{figure}[!htbp]
    \centering
    \includegraphics[width=0.90\textwidth]{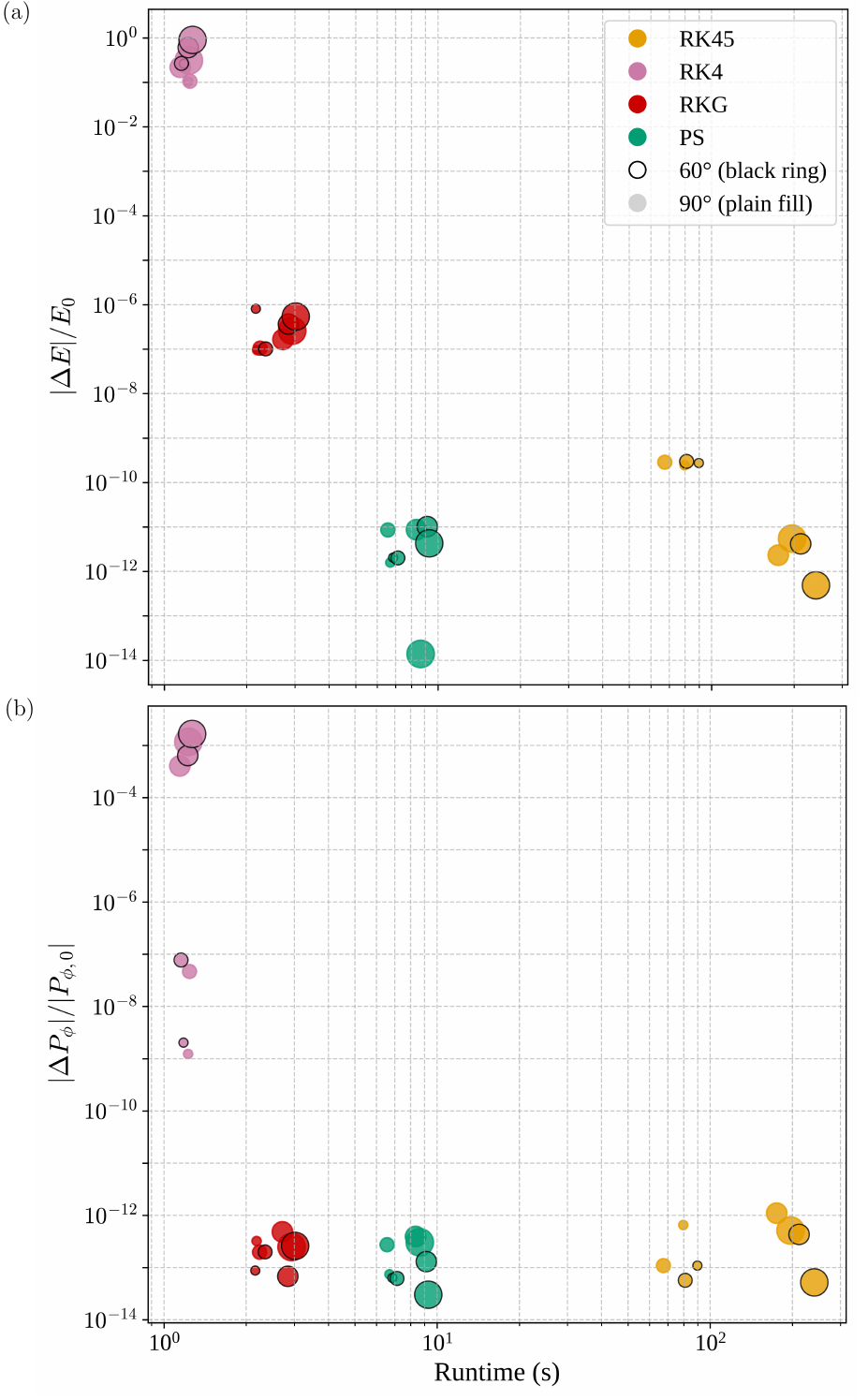}
    \caption{Same as Figure~\ref{FIG:bubbleplot_protons} but for electrons at four energies (50~keV, 1~MeV, 100~MeV, and 150~MeV), with black-ringed markers denoting the $60^{\circ}$ pitch angle runs and plain filled markers the $90^{\circ}$ runs. \remove[a]{Only the PS, RK4, and RK45 methods are shown; RKG is omitted due to its instability for electrons.}\note[R2]{Image regenerated}}
    \label{FIG:bubbleplot_electrons}
\end{figure}

Figure~\ref{FIG:bubbleplot_electrons} compares the trade-offs between invariant accuracy and computational runtime in our configuration: kinetic energy in panel~(a) and canonical momentum in panel~(b). Each marker represents the mean relative error computed over the final 1\% of the run. A complete tabulation of these metrics is provided in Appendix Table~\ref{tab:electron_energy_results}. 

As in the proton simulations, RK4 has the shortest runtimes at the expense of the largest errors in both invariants. \change[R2]{RK45 improves on RK4's accuracy but still conserves neither invariant well, and does so at the longest runtimes of any method; its relative kinetic energy error grows steadily, producing an increasing azimuthal phase error relative to the PS solution and, over longer integrations, the eventual loss of physically meaningful trajectories.}{RK45 has low errors in both invariants, with relative kinetic energy errors approximately two orders of magnitude larger than those of PS at lower energies and comparable at the highest energies, but at the longest runtimes of any method.} \add[a]{RKG has relative canonical momentum errors comparable to those of PS at shorter runtimes, but its relative kinetic energy errors are 4 to 7 orders of magnitude larger and, in some configurations, drift over the course of the integration, with an onset that varies between configurations.} By contrast, the PS method is the only method with \add[a]{consistently} low errors in both exact invariants, at runtimes longer than \change[a]{RK4's}{those of RK4 and RKG, at an identical fixed time step,} but well below those of RK45.

\clearpage

\subsection{A Case Study at a Target Accuracy}\label{SEC:Dip Discussion}
The previous simulations used identical time steps for the PS method and the fixed-step RK methods (RK4 and RKG), enabling direct, controlled comparisons\add[R2]{, while RK45 tolerances were set near their optimum for accuracy}. Under this configuration, \change[R2]{RK4}{fixed-step RK methods} generally had shorter runtimes at the expense of \remove[R2]{substantially} larger relative kinetic energy and canonical momentum errors, while the PS method generally produced significantly higher accuracy with correspondingly longer runtimes; the PS method typically required \change[a]{up to four}{several} times the computational runtime of \change[R2]{RK4 and runtimes comparable to RKG}{the fixed-step RK methods while RK45 runs were substantially longer than all other methods}, as summarized in Tables~\ref{tab:proton_energy_results} and~\ref{tab:electron_energy_results}. As noted in Section~\ref{SEC:intro}, the PS step size is linked to the convergence rate of the power series representation, enabling the use of larger time steps. This motivates an alternative comparison in which\change[R2]{PS, RK4, and RKG}{ all methods} are instead run to meet a target relative kinetic energy error with time steps chosen as large as practical \add[R2]{and less stringent tolerances used for RK45}, allowing computational efficiency to be assessed on the basis of a minimum accuracy requirement\remove[R2]{rather than a common step size}. Revisiting the 100~keV proton at a pitch angle of $30^\circ$, if a relative kinetic energy error below $10^{-5}$ is required over four drift periods, the PS method reaches this with a fixed time step of $\Delta\tau=54$, at which the mean truncation order rises to 22 (PS22); the RK4 method requires a much smaller step of $\Delta\tau=0.5$, \remove[R2]{and} the RKG method a step of $\Delta\tau=7.9$\add[R2]{, and RK45 requires relative and absolute tolerances of $10^{-13}$ and $10^{-15}$}. Under this comparison, the PS method completes the simulation \change[R2]{approximately ten times faster than RK4 and twice as fast as RKG}{several times faster than the explicit RK methods and comparable to RKG}. 

\begin{figure}[!htbp]
    \centering
    \includegraphics[width=\textwidth]{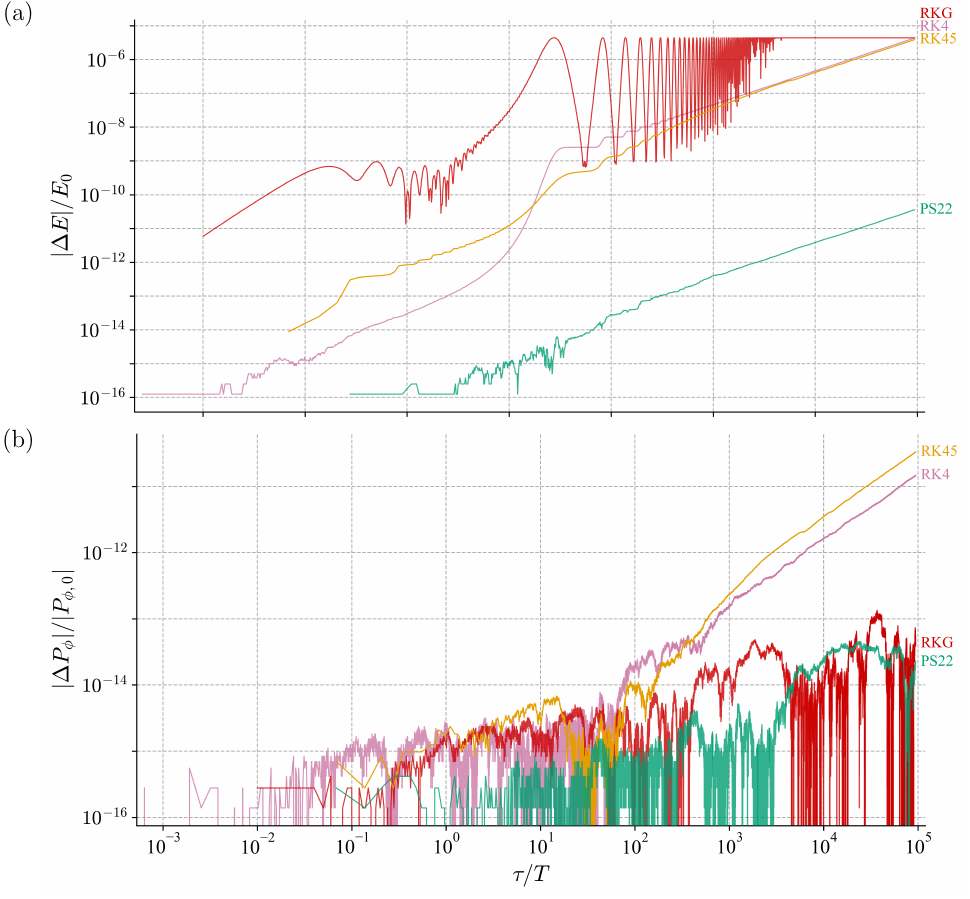}
    \caption{Relative errors of the exact invariants for the proton in Fig.~\ref{FIG:DipoleB_ex1_traj}: (a) kinetic energy and (b) canonical momentum. PS22 uses a fixed time step of $\Delta\tau = 54$, while RK4 uses a fixed time step of $\Delta\tau = 0.5$, \remove[R2]{and} RKG uses a fixed time step of $\Delta\tau = 7.9$\add[R2]{, and RK45 uses relative and absolute tolerances of $10^{-13}$ and $10^{-15}$, respectively}. All configurations were selected to reach a relative kinetic energy error below $10^{-5}$ over four drift periods while attempting to maximize computational efficiency.\note[R2]{Images were updated to include the updated RK45 analysis}}
    \label{fig:maxerror}
\end{figure}

The corresponding relative errors of the exact invariants are shown in Fig.~\ref{fig:maxerror}: kinetic energy in panel~(a) and canonical momentum in panel~(b). The first data points occur at different times due to the differing step sizes, resulting in an offset at the beginning of each curve. 

Despite its much larger time step, PS22 has a relative kinetic energy error 5 orders of magnitude lower than those of the RK runs, and a relative canonical momentum error 2 orders of magnitude lower than that of RK4 \add[R2]{and RK45,} and comparable to that of RKG. An analogous comparison for the electron simulations\remove[R2]{, using the PS and RK4 methods,} gives the same qualitative result.

Note that the 5 orders of magnitude difference in kinetic energy error between the PS method and the RK methods reflects a property of the PS method: increasing the step size raises the truncation order selected by the adaptive criterion while maintaining accuracy. Forcing the PS error up to the target accuracy required terminating the truncation early at an arbitrary fixed order, which increased the error without appreciably reducing runtime. Unlike the RK methods, whose error can be adjusted by the step size \add[R2]{or tolerance}, the PS method derives its efficiency gains from the larger time steps the series representation supports rather than from sacrificing accuracy. Conversely, forcing the RK methods to errors comparable to the PS method's required thousands of steps per gyroperiod\change[R2]{, with}{ or, for RK45, very stringent tolerances, resulting in} runtimes tens to hundreds of times slower than the PS method.

The results show that the PS method can meet the accuracy requirement with significantly larger time steps, allowing computational gains when stringent accuracy criteria define solution requirements for a given time interval. Although computational runtime was not the primary objective of this study, these findings indicate that, with appropriate parameter selection, the PS method can be not only accurate but also competitive or superior in efficiency for charged-particle applications. Adopting additional methods for adaptive order and step-size control could further improve these results. 

\clearpage

\section{Conclusions}\label{SEC:Conclusions}
The Parker--Sochacki power series (PS) method was shown to be an accurate and stable framework for long-time evaluation of charged-particle motion in a static magnetic dipole field. The results also establish that the PS algorithm is a viable method for evaluating the long-time behavior of the adiabatic invariants on timescales and in regimes where their conservation is not guaranteed by adiabatic theory \cite{DragtFinn1976}. The analysis reported here was motivated to address this problem, but the precision of this framework extends naturally to other space physics applications with similarly stringent accuracy requirements. These include cutoff-rigidity and magnetospheric access calculations, where trajectories must remain physically accurate over many gyroperiods \cite{Smart2000}, as well as relativistic proton scattering in the inner magnetosphere \cite{Meseguer2025} and the chaotically trapped particles around small magnetized bodies \cite{Oran2022}. 

%SMART2000: "The problem of determining which trajectories are allowed and which are forbidden is not as simple as it might initially seem. In the internal magnetic field representations and especially in the more complex magnetospheric fields, there is a set of low rigidity trajectories that have very long path length, consisting of many complex loops. Often for the sake of economy of computer resources, trajectory calculations are terminated after a large number of steps. This results in groups of indeterminate trajectories whose fate is not resolved. In Störmer theory there is a special set of trajectories which will have an arbitrary number of loops before reaching a final solution. In a simple dipole field, these low rigidity trajectories having many loops were generally forbidden."
%" The trajectory calculations are presumably as accurate as the geomagnetic field models utilized assuming that the numerical techniques yield an exact solution and the computers involved have sufficient numerical accuracy."
%"We have found a few noxious cases where, in complex particle trajectories near the cutoff rigidity, there were sudden, very small loops in the trajectory and the step size adjustment algorithm did not respond with sufficient agility to faithfully trace the trajectory."

These results also establish the PS method as an alternative to traditional RK-based methods. \change[R2]{Benchmarked at identical fixed time steps against established RK methods (RK4, RK45, and the symplectic RKG), the PS method preserved the exact invariants of the problem to high precision, with relative kinetic energy errors 4 to 13 orders of magnitude lower than those of the RK methods and canonical momentum errors 4 to 10 orders of magnitude lower than those of the explicit RK methods, comparable to, and typically within an order of magnitude below, those of RKG.}{Benchmarked against established RK methods (RK4 and the symplectic RKG at identical fixed time steps, and RK45 with adaptive steps at stringent tolerances), the PS method preserved the exact invariants of the problem to high precision, with relative kinetic energy errors approximately 10 or more orders of magnitude lower than those of RK4 and 4 or more orders of magnitude lower than those of RKG, and canonical momentum errors comparable to those of RKG. Only RK45 had comparable accuracy in both invariants, but with runtimes more than an order of magnitude longer.} A further insight from this study is that the symplectic structure of the RKG method does not guarantee bounded energy error over arbitrarily long integrations of the static magnetic dipole problem. Although its relative kinetic energy error is initially bounded, extended integrations reveal the eventual onset of error growth. \add[a]{A similar drift was observed in several electron configurations, with an onset that varies between configurations.} By contrast, the PS method's energy error grows approximately linearly but at a significantly smaller magnitude.

The PS method was compared directly with the RKG algorithm of \citeA{Yugo2007}. \change[a]{The key difference is that we find stability in Cartesian coordinates when the RKG algorithm failed and significantly lower energy error}{Unlike \mbox{\citeA{Yugo2007}}, we find both the PS method and our implementation of the RKG algorithm to be stable for electrons in Cartesian coordinates, with the PS method having significantly lower kinetic energy error} (\citeA{Yugo2007} did not provide an analysis of canonical momentum error). \citeA{Meseguer2025} used a 9th-order RK method modified with a projection over the invariant manifold. They reported energy and canonical momentum errors near machine precision over an integration time of 30~s using a constraint enforced at each integration step and using an adaptive step-size method, but the results for longer-time trajectory calculations were not reported.

As noted in Section~\ref{sect:implementation}, wall-clock computation times indicate the relative behavior of the methods rather than definitive rankings, and conclusions regarding computational efficiency depend on the basis of comparison. At identical fixed time steps, the PS method has longer runtimes than RK4\add[R2]{ and RKG} but with relative kinetic energy errors several orders of magnitude smaller. Because RK4 is the most computationally efficient fixed-step baseline considered here, step-matched comparisons understate the efficiency of the PS method when accuracy is the controlling constraint. When the methods are instead compared at a minimum target relative kinetic energy error, the PS method meets the target with substantially larger time steps; in representative proton and electron cases, it completes the simulations with runtimes \change[R2]{up to an order of magnitude shorter than those of the RK methods}{several times shorter than those of the explicit RK methods and comparable to those of RKG}. Finally, these comparisons may be sensitive to implementation choices such as iteration tolerances and starting values. We found that such information is under-reported in the available literature, requiring reconstruction from reported step--size or error levels.

\pagebreak
\appendix

\setcounter{equation}{0} 

\section{Examples - The Power Series Method}
\label{sect:examples}
In the PS formulation, all time-dependent variables are represented as power series expansions with recursively generated coefficients. We distinguish between two types of quantities: (i) dynamical variables, written without indexed subscripts (e.g., $x$, $v$), which are time-dependent variables governed directly by the ODE, and (ii) indexed quantities (e.g., $x_i$, $v_i$), which are constant coefficients of $t^i$ in the power series expansion that are solved for via recursion. Auxiliary variables are additional dynamical variables introduced to rewrite nonlinear functions into forms that allow for a recursion formula to be found \cite{Parker1996, Rudmin1998}.

We demonstrate the fundamentals of the method with three simple ODEs. Example I (Section \ref{Intro: ExI}) introduces the basic PS method for a 1-D ODE, $dy/dx=y$, where no auxiliary variables are needed, and it generates the exact series solution. Example II (Appendix \ref{Intro: ExII}) is a nonlinear 1-D ODE, $dy/dx=y^2$, and an auxiliary variable is used to linearize it and allow the exact solution to be determined with a recurrence relation. Example III (Appendix \ref{Intro: ExIII}) is a 2-D nonlinear ODE, and the PS method generates an approximate solution using an auxiliary variable and truncating the resulting power series. These examples establish the foundation for applying the PS method to the magnetic dipole field in Section~\ref{sec:dipole}.

\subsection{Example I - No Auxiliary Variables}\label{Intro: ExI}

To first demonstrate the PS method with no auxiliary variables, consider the ODE
    \begin{equation}\label{EQN: ExI dy/dx}
        \frac{dy}{dx}=y\quad\mbox{with}\quad y(0) =1
    \end{equation}
and express $y$ as a power series with unknown constant coefficients $y_i$,
    \begin{equation}\label{EQN: ExI y}
            y=\sum_{i=0}^\infty y_ix^i,
    \end{equation}
for which the derivative with respect to $x$ is
    \begin{equation}\label{EQN: ExI dy/dxi}
        \frac{dy}{dx}=\sum_{i=0}^\infty y_iix^{i-1}.
    \end{equation}
We see that $y_0=y(0)$ follows from Eq.~\eqref{EQN: ExI y}. Substituting Eqs. \eqref{EQN: ExI y} and \eqref{EQN: ExI dy/dxi} into Eq.~\eqref{EQN: ExI dy/dx} gives
    \begin{equation*}
            \sum_{i=0}^\infty y_iix^{i-1}=\sum_{i=0}^\infty y_ix^i.
    \end{equation*}
The recurrence relation is derived by first equating common powers of $x$ by reindexing the left-hand side using $i \to i+1$, giving
    \begin{equation*}
        \sum_{i=0}^\infty y_{i+1}(i+1)x^{i}=\sum_{i=0}^\infty y_ix^i.
    \end{equation*}
Hereafter, we will refer to this step simply as ``reindexing.''  Equating powers of $x$ gives
    \begin{equation}\label{EQN: ExI yi}
        y_{i+1} = \frac{y_i}{(i+1)}.
    \end{equation}
The first three coefficients are
    \begin{equation*}
        y_1=1,\quad\quad y_2=\frac{1}{2},\quad\mbox{and}\quad y_3=\frac{1}{6}.
    \end{equation*}
The general solution to Eq.~\eqref{EQN: ExI y}, using coefficients determined by Eq.~\eqref{EQN: ExI yi}, is thus
    \begin{equation*}
        y=1+x+\frac{x^2}{2!}+\frac{x^3}{3!}+\dots=e^x,
    \end{equation*}
which is the exact solution of Eq.~\eqref{EQN: ExI dy/dx}.

\subsection{Example II - Auxiliary Variable}\label{Intro: ExII}
The previous example demonstrates the use of a power series, but an auxiliary variable was not needed. Next, consider the nonlinear ODE
    \begin{equation}\label{EQN: ExII dydx}
        \frac{dy}{dx}=y^2\quad\mbox{with}\quad y(0) =1.
    \end{equation}
First, express $y$ as a power series,
    \begin{equation}\label{EQN: ExII y}
        y=\sum_{i=0}^\infty y_ix^i,   
    \end{equation}
where we see that $y_0=y(0)$. An auxiliary variable, 
    \begin{equation*}
           u \equiv y^2 = \left(\sum_{i=0}^\infty y_i x^i\right)^2,
    \end{equation*} 
can be simplified using the Cauchy product identity \cite{Knuth1997},
    \begin{equation}\label{eqn:cauchy}
        \left(\sum_{i=0}^\infty a_ix^i\right)\left(\sum_{j=0}^\infty b_jx^j\right)=\sum_{k=0}^\infty x^k\sum_{l=0}^{k} a_lb_{k-l},
    \end{equation}
giving
    \begin{equation}\label{EQN: ExII u}   
           u=\sum_{i=0}^\infty  x^i \left(\sum_{j=0}^i y_j y_{i-j}\right).
    \end{equation}      
Taking the derivative of Eq.~\eqref{EQN: ExII y} with respect to $x$ and reindexing gives  
    \begin{equation}\label{EQN: ExII dydxb}
        \frac{dy}{dx}=\sum_{i=0}^\infty (i+1)y_{i+1}x^i.
    \end{equation}
Equating Eq.~\eqref{EQN: ExII u} and Eq.~\eqref{EQN: ExII dydxb} gives
    \begin{equation*}
        \sum_{i=0}^\infty (i+1)y_{i+1}x^i=\sum_{i=0}^\infty x^i\left(\sum_{j=0}^iy_jy_{i-j}\right). 
        \end{equation*}
Equating powers of $x$ gives the recurrence relation,
    \begin{equation}
        y_{i+1}=\frac{1}{(i+1)}\left(\sum_{j=0}^iy_jy_{i-j}\right).\label{EQN: ExII yi}
    \end{equation}
The first three terms are
    \begin{align*}
            y_1=y_0y_0=&1,\\
            y_2=\frac{1}{2}\left(y_0y_1+y_1y_0\right)=&1,\\
     \intertext{and}    
            y_3=\frac{1}{3}\left(y_0y_2+y_1y_1+y_2y_0 \right)=&1.
    \end{align*}
By inspection, the general solution to Eq.~\eqref{EQN: ExII y}, using the coefficients determined by Eq. ~\eqref{EQN: ExII yi}, is 
    \begin{equation*}
        y=1+x+x^2+x^3+\dots=\frac{1}{1-x}\,,
    \end{equation*}
which is the exact solution of Eq.~\eqref{EQN: ExII dydx}. 

\subsection{Example III - ODE Solver}\label{Intro: ExIII}
The previous two examples demonstrated how the PS method constructs an exact power series solution and how auxiliary variables can be used to rewrite nonlinear terms to enable finding recurrence relations for the power series coefficients. While these problems could also be solved using the Picard method, they were selected to demonstrate the fundamental steps of the PS method. In these examples, the method directly produced closed-form power series exact solutions. However, most systems do not have a closed-form solution, and the Picard method becomes impractical due to the complexity of the algebraic expressions \cite{Ansari2024, Leal2015, Rudmin1998}. In this case, the PS method can be used as a numerical ODE solver by truncating the power series based on the number of meaningful contributing terms, defined as the maximum order for which the next term’s contribution to the sum is less than a defined tolerance, such as machine epsilon. This series is evaluated with a chosen time-step size of $\Delta t$ that determines how far in time the truncated series is used before it is reinitialized. When advanced iteratively, auxiliary variables can accumulate small numerical inconsistencies due to finite precision and truncation errors, leading to gradual deviation from their exact values. 

At the start of the next iteration, an additional stabilization strategy known as tethering \cite{Stewart2009} can be used to control the long-time accumulation of numerical error in the auxiliary variables. Tethering reinitializes the auxiliary variable at the end of each iteration to its analytical definition.

To demonstrate the setup of using the PS method for numerical ODE integration and tethering, we consider a system of two coupled ODEs corresponding to the one-dimensional motion of an object with quadratic drag:
    \begin{align}
            \frac{dx}{dt}=&\,v,\label{EQN: ExIII dx} \\
            \frac{dv}{dt}=&-v^2\nonumber.
    \end{align}
We define the auxiliary variable $u\equiv v^2$. The variables $x$, $v$, and $u$ as power series in $t$ are
    \begin{gather}
        x=\sum_{i=0}^\infty x_it^i \label{EQN: ExIII x},\\
        v=\sum_{i=0}^\infty v_it^i \label{EQN: ExIII v},\\
     \intertext{and}    
        u=\sum_{i=0}^\infty u_it^i  \label{EQN: ExIII u}.
    \end{gather}
From these power series, it follows that $x_0=x(0)$, $v_0=v(0)$, and $u_0=u(0)=v_0^2$. Taking the derivative of Eq.~\eqref{EQN: ExIII x} with respect to $t$ gives
    \begin{equation*}
        \frac{dx}{dt}=\sum_{i=0}^\infty ix_{i}t^{i-1}.
    \end{equation*}
Substituting these results and Eq.~\eqref{EQN: ExIII v} into Eq.~\eqref{EQN: ExIII dx}, and subsequently reindexing, gives  
    \begin{equation*}\label{EQN: ExIII sumvsumx}
        \sum_{i=0}^\infty (i+1)x_{i+1}t^{i}=\sum_{i=0}^\infty v_it^i  
    \end{equation*}
and equating powers of $t$ gives
    \begin{equation}\label{EQN: ExIII xi}
        x_{i+1}= \frac{v_i}{(i+1)}. 
    \end{equation}
Following the same process for Eq.~\eqref{EQN: ExIII v} gives
    \begin{equation}\label{EQN: ExIII vi}
        v_{i+1}= -\frac{u_i}{(i+1)}. 
    \end{equation}
For the final recurrence relation, note that
    \begin{align*}
        \frac{du}{dt}=2v\frac{dv}{dt}=-2vu.
    \end{align*}
Using this in the derivative of Eq.~\eqref{EQN: ExIII u} with respect to $t$ gives 
    \begin{equation*}
         \sum_{i=0}^\infty iu_it^{i-1}=\frac{du}{dt}=-2vu. 
    \end{equation*} 
Using Eqs. \eqref{EQN: ExIII v} and \eqref{EQN: ExIII u} gives
    \begin{equation*}
         \sum_{i=0}^\infty iu_it^{i-1}=-2\left( \sum_{i=0}^\infty v_it^i \right)\left( \sum_{i=0}^\infty u_it^i\right).
    \end{equation*}
Using the Cauchy product identity, Eq.~\eqref{eqn:cauchy}, gives
   \begin{equation*}
         \sum_{i=0}^\infty iu_it^{i-1}=-2\sum_{i=0}^\infty t^i\left(\sum_{j=0}^i v_ju_{i-j}\right).
    \end{equation*}
Reindexing the left-hand side and equating common powers of $t$ gives
    \begin{equation}\label{EQN: ExIII ui}
         u_{i+1}=-\frac{2}{i+1}\sum_{j=0}^iv_ju_{i-j}.
    \end{equation}
The recurrence relations, Eqs. \eqref{EQN: ExIII xi}, \eqref{EQN: ExIII vi}, and \eqref{EQN: ExIII ui}, can be substituted into Eqs. \eqref{EQN: ExIII x}, \eqref{EQN: ExIII v}, and \eqref{EQN: ExIII u}. Using the initial conditions and reindexing gives
    \begin{gather} \label{EQN: ExIII fulleqns}
        x(t)= x_0 + \sum_{i=0}^\infty\frac{v_i}{(i+1)} t^{i+1},\quad
        v(t)= v_0 - \sum_{i=0}^\infty \frac{u_i}{(i+1)}t^{i+1}, \quad\mbox{and}\quad        
        u(t)=u_0 - \sum_{i=0}^\infty \frac{2t^{i+1}}{i+1}\sum_{j=0}^iv_ju_{i-j}.
    \end{gather}

Although the relations in Eq.~\eqref{EQN: ExIII fulleqns} are expressed as an infinite series, a finite summation is needed for numerical calculations. The summations are terminated at an index denoted by the PS order, $M$. The PS order may be specified as a fixed value or set at each iteration by a criterion based on the significance of successive terms, for example, when additional contributions to the sum are less than a prescribed tolerance.

The recurrence relations in Eqs.~\eqref{EQN: ExIII xi}--\eqref{EQN: ExIII ui} were derived for the power series solution and were therefore written without iteration superscripts. When the PS method is used as a numerical ODE solver, these relations are applied at each time step $k$ to generate a new set of coefficients via
    \begin{gather}
        x_{i+1}^{(k)}=\frac{v_i^{(k)}}{i+1}, \label{eqn:EXIII xi ODE solver} \\
        v_{i+1}^{(k)}=-\frac{u_i^{(k)}}{i+1}, \label{eqn:EXIII vi ODE solver} \\
        u_{i+1}^{(k)}=-\frac{2}{i+1}\sum_{j=0}^i v_j^{(k)}u_{i-j}^{(k)}\label{eqn:EXIII ui ODE solver}\thinspace.
    \end{gather}
Similarly, the quantities $(x_{0},v_{0},u_{0})$ that previously denoted the values at $t=0$ now become iteration-dependent initial conditions $(x_{0}^{(k)},v_{0}^{(k)},u_{0}^{(k)})$, corresponding to the values at the beginning of each time step. These coefficients are then substituted into the truncated power series from Eq.~\eqref{EQN: ExIII fulleqns}, giving
       \begin{gather}
        x^{(k)}(\Delta t) \approx x_0^{(k)} + \sum_{i=0}^{M} \frac{v_i^{(k)}}{i+1}\,\Delta t^{i+1}, \label{eqn:EXIII x ODE solver}\\
        v^{(k)}(\Delta t) \approx v_0^{(k)} - \sum_{i=0}^{M} \frac{u_i^{(k)}}{i+1}\,\Delta t^{i+1}, \label{eqn:EXIII v ODE solver}\\
\intertext{and}
        u^{(k)}(\Delta t) \approx u_0^{(k)} - \sum_{i=0}^{M} \frac{2\Delta t^{i+1}}{i+1}
        \left(\sum_{j=0}^{i} v_j^{(k)}\,u_{i-j}^{(k)}\right).\label{eqn:EXIII u ODE solver}
    \end{gather}
The values $\big(x^{(k)}(\Delta t),\,v^{(k)}(\Delta t),\,u^{(k)}(\Delta t)\big)$ define the state at iteration $k$. These values are used as the initial conditions, $\big(x_0^{(k+1)},\, v_0^{(k+1)}, \,u_0^{(k+1)}\big)$, for the next iteration, $k+1$, thereby producing an iterative numerical ODE solver. Recall that a non-indexed symbol such as $x(t)$, $v(t)$, or $u(t)$  denotes a time-dependent variable in the ODE, whereas a subscripted quantity  such as $x_i$, $v_i$, or $u_i$ refers to the constant coefficient of $t^i$ in the power series expansion, and a superscript such as $x_0^{(k)}$, $v_0^{(k)}$, or $u_0^{(k)}$
denotes the value of the variable at iteration $k$.

After each iteration, the dynamical variables $x$ and $v$ are updated using recurrence relationships. The auxiliary variable $u=v^2$ can be updated in two ways. We can either use $u^{(k)}$ as the initial condition $u_0^{(k+1)}$ for the next iteration or we can use $u_0^{(k+1)}\simeq(v^{(k)}(\Delta t ))^2$, its analytical definition updated with the dynamical variable $v$. The latter choice, tethering, reinitializes the auxiliary variable to its exact value. To summarize, initial conditions when advancing from iteration $k$ to $k+1$ can be done in two ways:\\

\noindent
\begin{minipage}{0.48\textwidth}
\textbf{No Tethering}
    \begin{align*}
        x_0^{(k+1)} &= x^{(k)}\\
        v_0^{(k+1)} &= v^{(k)}\\
        u_0^{(k+1)} &= u^{(k)}
    \end{align*}
\emph{For each variable, the new initial condition is the last value from its truncated sum.\\}
\end{minipage}\hfill
\begin{minipage}{0.48\textwidth}
\textbf{Tethering}
    \begin{align*}
        x_0^{(k+1)} &= x^{(k)}\\
        v_0^{(k+1)} &= v^{(k)}\\
        u_0^{(k+1)} &= \left(v^{(k)}\right)^2
    \end{align*}
\emph{Auxiliary variable, $u$, reinitialized using its definition, $u=v^2$. For dynamical variables, the new initial condition is the last value from the truncated sum.}
\end{minipage}\\

\vspace{0.5cm}
To demonstrate the step-wise use of the recurrence relations, consider iteration $k=1$ and let $M=1$ and $\Delta t=1$.  From Eqs. \eqref{eqn:EXIII xi ODE solver}-\eqref{eqn:EXIII ui ODE solver}, the coefficient updates become
    \begin{gather*}
        x_{1}^{(1)}=v_0^{(1)}, \\
        v_{1}^{(1)}=-u_0^{(1)},  \\
\intertext{and}
        u_{1}^{(1)}=-2v_0^{(1)}u_{0}^{(1)}.
    \end{gather*}
Applying these to Eqs. \eqref{eqn:EXIII x ODE solver}-\eqref{eqn:EXIII u ODE solver}, we obtain
    \begin{gather*}
        x^{(1)}(1) \approx x_0^{(1)} + v_0^{(1)}
        - \tfrac12\big(v_0^{(1)}\big)^{2},\\
        v^{(1)}(1) \approx v_0^{(1)} - u_0^{(1)} - \tfrac12 u_1^{(1)},\\
\intertext{and}
        u^{(1)}(1) \approx  u_0^{(1)} - 2v_0^{(1)}u_0^{(1)} - v_0^{(1)}u_1^{(1)} - v_1^{(1)}u_0^{(1)}.
    \end{gather*}
This completes the first PS iteration and provides the values that become the initial conditions for iteration $k=2$ under either tethered, $u_0^{(2)}=(v^{(1)})^2$, or untethered, $u_0^{(2)}=u^{(1)}$, updating of the auxiliary variable.

\section{Supplemental Tables}\label{appendix:tables}
\renewcommand{\thetable}{B.\arabic{table}}
\sisetup{exponent-product=\cdot}

\begin{table}[H]
    \setlength{\extrarowheight}{2pt}
    \centering
    \caption{Proton simulation of mean and maximum relative kinetic energy and canonical momentum error in a magnetic dipole field with runtime for each method. Red text highlights method instability\remove[R2]{, and N/A indicates omitted method. In one run, the RK45 method was excluded from the analyses due to computational stalling or solver instability, which rendered completion untenable}. Mean and maximum values were computed over the final 1\% of the run, $\tau/T=1.5\times 10^5$ equatorial gyroperiods with a fixed time step of $\Delta\tau=12.1$ \add[R2]{(RK45 uses adaptive steps with the tolerances given in Section~\ref{sect:implementation}),} initialized at $L=5$. \note[R2]{Table regenerated with RK45 method now using a near-optimal tolerance for accuracy}}
    \label{tab:proton_energy_results}
    \setlength{\tabcolsep}{4pt}  
\begin{tabular}{|c|c|c|c|c|c|c|}
\hline
\textbf{Energy} & $\gamma$ & \textbf{Pitch} & \textbf{Method} & $|\Delta E|/E_0$ \textbf{(mean / max)} & $|\Delta P_\phi|/P_{\phi_0}$ \textbf{(mean / max)} & \textbf{Runtime (s)} \\
\hline
\multirow{8}{*}{10 keV} & \multirow{8}{*}{1.00} & \multirow{4}{*}{90$^\circ$} & RK45 & $9.99 \cdot 10^{-11}$ / $1.00 \cdot 10^{-10}$ & $3.38 \cdot 10^{-13}$ / $4.91 \cdot 10^{-13}$ & 81.6 \\
\cline{4-7}
 &  &  & RK4 & $1.01 \cdot 10^{-1}$ / $1.02 \cdot 10^{-1}$ & $4.21 \cdot 10^{-7}$ / $4.23 \cdot 10^{-7}$ & 1.3 \\
\cline{4-7}
 &  &  & RKG & $3.87 \cdot 10^{-8}$ / $4.52 \cdot 10^{-8}$ & $1.79 \cdot 10^{-13}$ / $2.00 \cdot 10^{-13}$ & 2.5 \\
\cline{4-7}
 &  &  & PS11 & $1.74 \cdot 10^{-12}$ / $1.79 \cdot 10^{-12}$ & $2.28 \cdot 10^{-14}$ / $3.39 \cdot 10^{-14}$ & 6.6 \\
\cline{3-7}
 &  & \multirow{4}{*}{30$^\circ$} & RK45 & $1.99 \cdot 10^{-10}$ / $2.00 \cdot 10^{-10}$ & $5.60 \cdot 10^{-13}$ / $6.16 \cdot 10^{-13}$ & 172.0 \\
\cline{4-7}
 &  &  & RK4 & \textcolor{red}{$2.81 \cdot 10^{6}$ / $2.81 \cdot 10^{6}$} & \textcolor{red}{$2.29 \cdot 10^{-2}$ / $2.29 \cdot 10^{-2}$} & \textcolor{red}{1.3} \\
\cline{4-7}
 &  &  & RKG & $6.57 \cdot 10^{-6}$ / $2.90 \cdot 10^{-5}$ & $5.63 \cdot 10^{-13}$ / $5.73 \cdot 10^{-13}$ & 3.2 \\
\cline{4-7}
 &  &  & PS13 & $1.38 \cdot 10^{-11}$ / $1.39 \cdot 10^{-11}$ & $1.21 \cdot 10^{-13}$ / $1.28 \cdot 10^{-13}$ & 8.5 \\
\hline
\multirow{8}{*}{100 keV} & \multirow{8}{*}{1.00} & \multirow{4}{*}{90$^\circ$} & RK45 & $3.16 \cdot 10^{-11}$ / $3.19 \cdot 10^{-11}$ & $1.08 \cdot 10^{-12}$ / $1.17 \cdot 10^{-12}$ & 108.1 \\
\cline{4-7}
 &  &  & RK4 & $9.50 \cdot 10^{-2}$ / $9.55 \cdot 10^{-2}$ & $4.00 \cdot 10^{-6}$ / $4.02 \cdot 10^{-6}$ & 1.2 \\
\cline{4-7}
 &  &  & RKG & $4.94 \cdot 10^{-8}$ / $6.84 \cdot 10^{-8}$ & $5.61 \cdot 10^{-13}$ / $5.69 \cdot 10^{-13}$ & 2.4 \\
\cline{4-7}
 &  &  & PS11 & $3.91 \cdot 10^{-12}$ / $3.93 \cdot 10^{-12}$ & $2.73 \cdot 10^{-13}$ / $2.82 \cdot 10^{-13}$ & 6.6 \\
\cline{3-7}
 &  & \multirow{4}{*}{30$^\circ$} & RK45 & $5.95 \cdot 10^{-11}$ / $5.98 \cdot 10^{-11}$ & $1.01 \cdot 10^{-12}$ / $1.06 \cdot 10^{-12}$ & 202.1 \\
\cline{4-7}
 &  &  & RK4 & \textcolor{red}{$3.62 \cdot 10^{5}$ / $3.62 \cdot 10^{5}$} & \textcolor{red}{$4.22 \cdot 10^{-1}$ / $4.22 \cdot 10^{-1}$} & \textcolor{red}{1.2} \\
\cline{4-7}
 &  &  & RKG & $5.56 \cdot 10^{-6}$ / $2.50 \cdot 10^{-5}$ & $8.13 \cdot 10^{-13}$ / $8.32 \cdot 10^{-13}$ & 2.9 \\
\cline{4-7}
 &  &  & PS13 & $1.43 \cdot 10^{-11}$ / $1.44 \cdot 10^{-11}$ & $4.95 \cdot 10^{-14}$ / $5.89 \cdot 10^{-14}$ & 8.5 \\
\hline
\multirow{8}{*}{1 MeV} & \multirow{8}{*}{1.00} & \multirow{4}{*}{90$^\circ$} & RK45 & $1.02 \cdot 10^{-11}$ / $1.04 \cdot 10^{-11}$ & $1.13 \cdot 10^{-12}$ / $1.24 \cdot 10^{-12}$ & 163.9 \\
\cline{4-7}
 &  &  & RK4 & $7.76 \cdot 10^{-2}$ / $7.80 \cdot 10^{-2}$ & $3.43 \cdot 10^{-5}$ / $3.44 \cdot 10^{-5}$ & 1.2 \\
\cline{4-7}
 &  &  & RKG & $8.13 \cdot 10^{-8}$ / $1.28 \cdot 10^{-7}$ & $1.72 \cdot 10^{-12}$ / $1.73 \cdot 10^{-12}$ & 2.4 \\
\cline{4-7}
 &  &  & PS12 & $3.11 \cdot 10^{-12}$ / $3.12 \cdot 10^{-12}$ & $1.17 \cdot 10^{-13}$ / $1.27 \cdot 10^{-13}$ & 7.0 \\
\cline{3-7}
 &  & \multirow{4}{*}{30$^\circ$} & RK45 & $1.68 \cdot 10^{-11}$ / $1.70 \cdot 10^{-11}$ & $8.68 \cdot 10^{-13}$ / $9.26 \cdot 10^{-13}$ & 243.4 \\
\cline{4-7}
 &  &  & RK4 & \textcolor{red}{$4.26 \cdot 10^{4}$ / $4.26 \cdot 10^{4}$} & \textcolor{red}{$9.94 \cdot 10^{-2}$ / $9.94 \cdot 10^{-2}$} & \textcolor{red}{1.2} \\
\cline{4-7}
 &  &  & RKG & $3.40 \cdot 10^{-6}$ / $1.58 \cdot 10^{-5}$ & $1.45 \cdot 10^{-12}$ / $1.47 \cdot 10^{-12}$ & 3.0 \\
\cline{4-7}
 &  &  & PS13 & $1.01 \cdot 10^{-11}$ / $1.02 \cdot 10^{-11}$ & $1.84 \cdot 10^{-14}$ / $2.74 \cdot 10^{-14}$ & 8.7 \\
\hline
\multirow{8}{*}{10 MeV} & \multirow{8}{*}{1.01} & \multirow{4}{*}{90$^\circ$} & RK45 & $9.79 \cdot 10^{-13}$ / $1.12 \cdot 10^{-12}$ & $1.73 \cdot 10^{-13}$ / $2.82 \cdot 10^{-13}$ & 133.3 \\
\cline{4-7}
 &  &  & RK4 & $4.45 \cdot 10^{-2}$ / $4.48 \cdot 10^{-2}$ & $2.22 \cdot 10^{-4}$ / $2.23 \cdot 10^{-4}$ & 1.2 \\
\cline{4-7}
 &  &  & RKG & $2.03 \cdot 10^{-7}$ / $2.83 \cdot 10^{-7}$ & $8.83 \cdot 10^{-13}$ / $8.96 \cdot 10^{-13}$ & 2.5 \\
\cline{4-7}
 &  &  & PS12 & $7.23 \cdot 10^{-12}$ / $7.27 \cdot 10^{-12}$ & $6.24 \cdot 10^{-14}$ / $7.82 \cdot 10^{-14}$ & 7.6 \\
\cline{3-7}
 &  & \multirow{4}{*}{30$^\circ$} & RK45 & $1.14 \cdot 10^{-13}$ / $2.47 \cdot 10^{-13}$ & $1.68 \cdot 10^{-12}$ / $1.77 \cdot 10^{-12}$ & 248.7 \\
\cline{4-7}
 &  &  & RK4 & $5.65 \cdot 10^{-1}$ / $5.66 \cdot 10^{-1}$ & $8.46 \cdot 10^{-4}$ / $8.49 \cdot 10^{-4}$ & 1.2 \\
\cline{4-7}
 &  &  & RKG & $8.65 \cdot 10^{-7}$ / $4.63 \cdot 10^{-6}$ & $3.93 \cdot 10^{-12}$ / $3.96 \cdot 10^{-12}$ & 3.0 \\
\cline{4-7}
 &  &  & PS13 & $4.55 \cdot 10^{-12}$ / $4.57 \cdot 10^{-12}$ & $3.49 \cdot 10^{-13}$ / $3.56 \cdot 10^{-13}$ & 8.9 \\
\hline
\end{tabular}
\end{table}

\renewcommand{\thetable}{B.\arabic{table}}
\begin{table}[h]
\setlength{\extrarowheight}{2pt}
\centering
\caption{Electron simulation of mean and maximum relative kinetic energy error and relative canonical momentum error in a magnetic dipole field with runtime for each method. \remove[a]{N/A indicates omitted method. In one run, the RK45 method was excluded from the analyses due to computational stalling or solver instability, which rendered completion untenable. The remaining omitted runs were the RKG method, which was unstable for electrons in Cartesian coordinates.} Mean and maximum values were computed over the final 1\% of the run, $\tau/T= 1.5\times 10^5$ equatorial gyroperiods with a fixed time step of $\Delta\tau=12.1$ \add[R2]{(RK45 uses adaptive steps with the tolerances given in Section~\ref{sect:implementation}),} initialized at $L=5$. \note[R2]{Table regenerated with RK45 method now using a near-optimal tolerance for accuracy. Additionally RKG values are no longer N/A after correcting a sign error in our implementation.}} 
\label{tab:electron_energy_results}
\setlength{\tabcolsep}{4pt}  
\begin{tabular}{|c|c|c|c|c|c|c|}
\hline
\textbf{Energy} & $\gamma$ & \textbf{Pitch} & \textbf{Method} & $|\Delta E|/E_0$ \textbf{(mean / max)} & $|\Delta P_\phi|/P_{\phi_0}$ \textbf{(mean / max)} & \textbf{Runtime (s)} \\
\hline
\multirow{8}{*}{50 keV} & \multirow{8}{*}{1.10} & \multirow{4}{*}{90$^\circ$} & RK45 & $2.43 \cdot 10^{-10}$ / $2.45 \cdot 10^{-10}$ & $6.53 \cdot 10^{-13}$ / $7.18 \cdot 10^{-13}$ & 79.7 \\
\cline{4-7}
 &  &  & RK4 & $1.05 \cdot 10^{-1}$ / $1.05 \cdot 10^{-1}$ & $1.23 \cdot 10^{-9}$ / $1.24 \cdot 10^{-9}$ & 1.2 \\
\cline{4-7}
 &  &  & RKG & $9.51 \cdot 10^{-8}$ / $9.61 \cdot 10^{-8}$ & $3.22 \cdot 10^{-13}$ / $3.37 \cdot 10^{-13}$ & 2.2 \\
\cline{4-7}
 &  &  & PS11 & $1.58 \cdot 10^{-12}$ / $1.62 \cdot 10^{-12}$ & $7.44 \cdot 10^{-14}$ / $9.23 \cdot 10^{-14}$ & 6.7 \\
\cline{3-7}
 &  & \multirow{4}{*}{60$^\circ$} & RK45 & $2.74 \cdot 10^{-10}$ / $2.76 \cdot 10^{-10}$ & $1.09 \cdot 10^{-13}$ / $1.56 \cdot 10^{-13}$ & 89.8 \\
\cline{4-7}
 &  &  & RK4 & $2.63 \cdot 10^{-1}$ / $2.65 \cdot 10^{-1}$ & $2.03 \cdot 10^{-9}$ / $2.05 \cdot 10^{-9}$ & 1.2 \\
\cline{4-7}
 &  &  & RKG & $8.08 \cdot 10^{-7}$ / $9.06 \cdot 10^{-7}$ & $8.76 \cdot 10^{-14}$ / $1.00 \cdot 10^{-13}$ & 2.2 \\
\cline{4-7}
 &  &  & PS11 & $2.05 \cdot 10^{-12}$ / $2.06 \cdot 10^{-12}$ & $6.34 \cdot 10^{-14}$ / $7.13 \cdot 10^{-14}$ & 6.9 \\
\hline
\multirow{8}{*}{1 MeV} & \multirow{8}{*}{2.96} & \multirow{4}{*}{90$^\circ$} & RK45 & $2.86 \cdot 10^{-10}$ / $2.88 \cdot 10^{-10}$ & $1.09 \cdot 10^{-13}$ / $1.66 \cdot 10^{-13}$ & 67.4 \\
\cline{4-7}
 &  &  & RK4 & $1.06 \cdot 10^{-1}$ / $1.06 \cdot 10^{-1}$ & $4.68 \cdot 10^{-8}$ / $4.70 \cdot 10^{-8}$ & 1.2 \\
\cline{4-7}
 &  &  & RKG & $1.05 \cdot 10^{-7}$ / $1.08 \cdot 10^{-7}$ & $1.97 \cdot 10^{-13}$ / $2.26 \cdot 10^{-13}$ & 2.2 \\
\cline{4-7}
 &  &  & PS11 & $8.55 \cdot 10^{-12}$ / $8.58 \cdot 10^{-12}$ & $2.74 \cdot 10^{-13}$ / $2.80 \cdot 10^{-13}$ & 6.6 \\
\cline{3-7}
 &  & \multirow{4}{*}{60$^\circ$} & RK45 & $2.98 \cdot 10^{-10}$ / $2.99 \cdot 10^{-10}$ & $5.70 \cdot 10^{-14}$ / $1.45 \cdot 10^{-13}$ & 81.0 \\
\cline{4-7}
 &  &  & RK4 & $2.67 \cdot 10^{-1}$ / $2.68 \cdot 10^{-1}$ & $7.74 \cdot 10^{-8}$ / $7.77 \cdot 10^{-8}$ & 1.2 \\
\cline{4-7}
 &  &  & RKG & $1.01 \cdot 10^{-7}$ / $1.82 \cdot 10^{-7}$ & $1.99 \cdot 10^{-13}$ / $2.14 \cdot 10^{-13}$ & 2.3 \\
\cline{4-7}
 &  &  & PS11 & $2.01 \cdot 10^{-12}$ / $2.03 \cdot 10^{-12}$ & $6.19 \cdot 10^{-14}$ / $7.14 \cdot 10^{-14}$ & 7.1 \\
\hline
\multirow{8}{*}{100 MeV} & \multirow{8}{*}{197} & \multirow{4}{*}{90$^\circ$} & RK45 & $2.32 \cdot 10^{-12}$ / $2.45 \cdot 10^{-12}$ & $1.10 \cdot 10^{-12}$ / $1.16 \cdot 10^{-12}$ & 175.2 \\
\cline{4-7}
 &  &  & RK4 & $2.18 \cdot 10^{-1}$ / $2.19 \cdot 10^{-1}$ & $4.04 \cdot 10^{-4}$ / $4.05 \cdot 10^{-4}$ & 1.1 \\
\cline{4-7}
 &  &  & RKG & $1.65 \cdot 10^{-7}$ / $4.91 \cdot 10^{-7}$ & $4.83 \cdot 10^{-13}$ / $4.93 \cdot 10^{-13}$ & 2.7 \\
\cline{4-7}
 &  &  & PS12 & $8.69 \cdot 10^{-12}$ / $8.76 \cdot 10^{-12}$ & $3.93 \cdot 10^{-13}$ / $4.07 \cdot 10^{-13}$ & 8.3 \\
\cline{3-7}
 &  & \multirow{4}{*}{60$^\circ$} & RK45 & $4.16 \cdot 10^{-12}$ / $4.27 \cdot 10^{-12}$ & $4.31 \cdot 10^{-13}$ / $5.51 \cdot 10^{-13}$ & 211.5 \\
\cline{4-7}
 &  &  & RK4 & $6.06 \cdot 10^{-1}$ / $6.08 \cdot 10^{-1}$ & $6.38 \cdot 10^{-4}$ / $6.41 \cdot 10^{-4}$ & 1.2 \\
\cline{4-7}
 &  &  & RKG & $3.65 \cdot 10^{-7}$ / $1.35 \cdot 10^{-6}$ & $6.80 \cdot 10^{-14}$ / $9.04 \cdot 10^{-14}$ & 2.8 \\
\cline{4-7}
 &  &  & PS13 & $1.02 \cdot 10^{-11}$ / $1.02 \cdot 10^{-11}$ & $1.30 \cdot 10^{-13}$ / $1.41 \cdot 10^{-13}$ & 9.1 \\
\hline
\multirow{8}{*}{150 MeV} & \multirow{8}{*}{295} & \multirow{4}{*}{90$^\circ$} & RK45 & $5.47 \cdot 10^{-12}$ / $5.64 \cdot 10^{-12}$ & $5.12 \cdot 10^{-13}$ / $6.19 \cdot 10^{-13}$ & 196.8 \\
\cline{4-7}
 &  &  & RK4 & $3.10 \cdot 10^{-1}$ / $3.11 \cdot 10^{-1}$ & $1.17 \cdot 10^{-3}$ / $1.17 \cdot 10^{-3}$ & 1.2 \\
\cline{4-7}
 &  &  & RKG & $2.60 \cdot 10^{-7}$ / $9.46 \cdot 10^{-7}$ & $2.48 \cdot 10^{-13}$ / $2.57 \cdot 10^{-13}$ & 2.9 \\
\cline{4-7}
 &  &  & PS13 & $1.39 \cdot 10^{-14}$ / $3.47 \cdot 10^{-14}$ & $3.05 \cdot 10^{-13}$ / $3.15 \cdot 10^{-13}$ & 8.6 \\
\cline{3-7}
 &  & \multirow{4}{*}{60$^\circ$} & RK45 & $4.87 \cdot 10^{-13}$ / $6.59 \cdot 10^{-13}$ & $5.20 \cdot 10^{-14}$ / $1.35 \cdot 10^{-13}$ & 240.5 \\
\cline{4-7}
 &  &  & RK4 & $9.00 \cdot 10^{-1}$ / $9.02 \cdot 10^{-1}$ & $1.65 \cdot 10^{-3}$ / $1.65 \cdot 10^{-3}$ & 1.3 \\
\cline{4-7}
 &  &  & RKG & $5.41 \cdot 10^{-7}$ / $2.42 \cdot 10^{-6}$ & $2.60 \cdot 10^{-13}$ / $2.71 \cdot 10^{-13}$ & 3.0 \\
\cline{4-7}
 &  &  & PS13 & $4.28 \cdot 10^{-12}$ / $4.29 \cdot 10^{-12}$ & $3.02 \cdot 10^{-14}$ / $4.15 \cdot 10^{-14}$ & 9.3 \\
\hline
\end{tabular}
\end{table}

\clearpage

\section*{Open Research}
The PS and RK integrators and the analysis scripts used to generate all results in this study (version v1.\change[a]{0}{1}.0, released under a Creative Commons Attribution 4.0 International license) are available in the Zenodo repository at \change[a]{\mbox{\url{https://doi.org/10.5281/zenodo.21026049}}}{\mbox{\url{https://doi.org/10.5281/zenodo.23504051}}}\cite{Jiles_PS_code}. No previously published or third-party datasets were used; all data presented in this work are produced by the archived software.

\section*{\add[a]{Use of Generative AI}}
\add[a]{Generative AI tools were used for late-stage revisions of text and code. Claude was used to (1) modify the style and annotations of existing plotting code and (2) refactor and reorganize existing code without modifying its functionality or algorithms; the refactored code is the version archived in the repository described in the Open Research Section, and all results were regenerated with it. Grammarly was used for minor revisions of grammar, spelling, punctuation, and diction in text written and revised by the authors. The authors take full responsibility for the content of the manuscript and the archived code.}\note[a]{moved from the Numerical Methods section}

\section*{Conflict of Interest Declaration}
The authors declare there are no conflicts of interest for this manuscript.

\acknowledgments
This work was informed by unpublished notes by Reginald Ford (2015), which provided initial context for the problem. We also thank Roger Thelwell and James Sochacki for feedback on this manuscript.

\bibliography{ref}
\end{document}